\documentclass[a4paper]{jpconf}
\usepackage{graphicx}

\usepackage{amsmath}
\usepackage{amssymb}
\usepackage{pdflscape}
\usepackage[titletoc]{appendix}

\usepackage{color}
\usepackage{latexsym}
\begin{document}

\title{Application of the admitted Lie group of the 
\\
classical Boltzmann equation to classification of the 
\\
Boltzmann equation
with a source term}
\author{Adisak Karnbanjong$^{a}$, Amornrat Suriyawichitseranee$^{a}$, 
\\
Yurii N. Grigoriev$^{b}$, Sergey V. Meleshko$^{a}$}
\address{$^{a}$ School of Mathematics, Institute of Science, Suranaree University
of Technology, \\
 Nakhon Ratchasima, 30000, Thailand \\
 $^{b}$ Institute of Computational Technology, Novosibirsk 630090, Russia
 }
\ead{kadisak@g-mail.wu.ac.th, amornratjulie@gmail.com, grigor@ict.nsc.ru, sergey@math.sut.ac.th}

\begin{abstract}
	The classical Boltzmann equation is an integro-differential equation which describes the time evolution of rarefied gas in terms of a molecular distribution function. For some kinetic problems where it is necessary to add in the Boltzmann equation a source term depending on the independent and dependent variables. This paper is devoted to applying preliminary group classification to the Boltzmann equation with a source function by using the Lie group $L_{11}$ admitted by the classical Boltzmann equation.
	
	The developed strategy for deriving determining equation of an integro-differential equation with a source (in general form) using a known Lie group admitted by the corresponding equation without the source is applied to the Boltzmann equation with a source. Solving the determining equation for the source function for each subalgebra of the optimal system of subalgebras of the Lie algebra $L_{11}$, a preliminary group classification of the Boltzmann equation with respect to the source function is obtained.
	
	Furthermore, representations of invariant solutions of the Boltzmann equation with a source are presented. The reduced equations are also shown for some representations of invariant solutions.
	
\end{abstract}


\section{Introduction}

\subsection{The full Boltzmann equation}
Behavior of molecules of a rarefied gas is described by the Boltzmann equation in terms of the distribution function $f$ of molecules of the gas
\begin{equation}\label{eq:fullBoltzmannEqWithSource}
	f_{t} + \textbf{v}\cdot \nabla_{\textbf{x}}f = J(f,f),
\end{equation}
where $J(f,f)$ is the collision integral
\begin{align}\label{collisionIntegral}
	J(f,f)=&\int\limits_{\mathbb{R}^{3}}\int\limits_{S^{2}}B(g,\theta_{1})(f^{*}f_{1}^{*}-ff_{1})\,d\textbf{n}\,d\textbf{w}.
\end{align}
Here $t$ is time, $\nabla_{\textbf{x}}$ is the gradient with respect to the space variable $\textbf{x}\in \mathbb{R}^{3}$,
$B(g,\theta_{1})$ is the collision scattering function, $f=f(\textbf{x},\textbf{v},t)$,~$f_{1}=f(\textbf{x},\textbf{w},t)$,~$f^{*}=f(\textbf{x},\textbf{v}^{*},t)$,~$f_{1}^{*}=f(\textbf{x},\textbf{w}^{*},t)$,~$\textbf{g}=\textbf{v}-\textbf{w}$ is the relative velocity of the colliding particles, $g=\|\textbf{g}\|_{2}$,~$\textbf{w}=(u_{1},v_{1},w_{1})\in\mathbb{R}^{3}$,~$d\textbf{w}=du_{1}dv_{1}dw_{1}$ is a volume element in $\mathbb{R}^{3}$, and $\textbf{v}, \textbf{w}$ are the pre-collision velocities of two particles having the post-collision velocities $\textbf{v}^{*}, \textbf{w}^{*}$. The velocities $\textbf{v}^{*}$ and $\textbf{w}^{*}$ are determined by following formulae
\begin{align*}
	\textbf{v}^{*}=\frac{1}{2}(\textbf{v}+\textbf{w}+g\textbf{n}),~
	\textbf{w}^{*}=\frac{1}{2}(\textbf{v}+\textbf{w}-g\textbf{n}),
\end{align*}
where $\textbf{n}$ is a unit vector varying on the unit sphere $S^{2}=\{\textbf{n}\in\mathbb{R}^{3}| ~\|\textbf{n}\|_{2}=1\}$ such as $\textbf{n}\equiv(n_{1},n_{2},n_{3})=(\sin{\theta_{1}}\cos{\epsilon},\sin{\theta_{1}}\sin{\epsilon},\cos{\theta_{1}})$, $\epsilon,~\theta_{1}$ are the angles identifying a point on the sphere in spherical coordinates, and $d\textbf{n}$ is a surface element of unit sphere in $\mathbb{R}^{3}$, i.e., $d\textbf{n}=\sin{\theta_{1}}d\theta_{1}d\epsilon$, where $\theta_{1}\in\left[0,\pi\right],~\epsilon\in\left[0,2\pi\right]$.

\subsection{Group analysis of the classical Boltzmann equation}

The presence of the complicated collision integral is one of the main difficulties for finding solutions of the Boltzmann equation.
Researchers have tried to find solutions of the Boltzmann equation with simplified collision integral.

In 1975, A. V. Bobylev found a particular solution of the spatially homogeneous Boltzmann equation with Maxwell molecules by applying the Fourier transform and representing the solution in a special form \cite{bk:Bobylev[1975b]}. Later, in 1976, M. Krook and T. T. Wu obtained the same solution by using the moment generating function \cite{bk:KrookWu[1976]}. This solution  (\textit{BKW solution})  was obtained by representing it in a special form \cite{bk:Bobylev[1975b]} with a reduced number of independent variables.

One of the methods which allows finding solutions with a reduced number of independent variables is the group analysis method. The first application of group analysis to integro-differential kinetic equations was in \cite{bk:Taranov[1976]}, where the Vlasov equations of collisionless plasma were reduced to a system of equations for moments, and the classical group analysis method was applied to a subsystem  which consists of finite number of partial differential equations. Then the number of studied equations was extended to infinity. The method \cite{bk:Taranov[1976]} was applied in \cite{bk:BunKras[1982],bk:BunKras[1983]} to the Bhatnagar-Gross-Krook (BGK) equation, which is a simplified model of the Boltzmann equation. In particular, in \cite{bk:BunKras[1982],bk:BunKras[1983]} it was found that the admitted Lie group of the BGK equation corresponds to the Lie algebra $L_{11}$ spanned by the basis generators:
\begin{align}\label{basisGeneratorsOfL11}
	X_{1} = \partial _x,~ X_{2} = \partial _y,~ X_{3} = \partial _z,~ X_{4} = t \partial _x + \partial _u,~ X_{5} = t \partial _y + \partial _v,~ \nonumber\\
	X_{6} = t \partial _z + \partial _w,~ X_{7} = y \partial _z - z \partial _y + v \partial _w - w \partial _v,~ X_{8} = z \partial _x - x \partial _z + w \partial _u - u \partial _w,\nonumber\\
	~ X_{9} = x \partial _y - y \partial _x + u \partial _v - v \partial _u,~ X_{10} = \partial _t,~ X_{11} = t \partial _t + x \partial _x + y \partial _y + z \partial _z - f \partial _f.
\end{align}
It should be noted that in \cite{bk:GrigMel[1999],bk:GrigorievMeleshko[2001]} it is shown that the Lie algebra (\ref{basisGeneratorsOfL11}) is isomorphic to the Lie algebra admitted by the gas dynamics equations \cite{bk:Ovsiannikov[1978],bk:Ovsiannikov[1994b]}.

In \cite{bk:GrigMel[1986],bk:GrigMel[1987]}, it was developed a direct method for applying group analysis to integro-differential equations. In \cite{bk:BobylevDorodnitsyn[2009]} the direct method was applied to the Boltzmann equation. More details can be found in \cite{bk:GrigorievIbragimovKovalevMeleshko2010}.

In order to model some physical situations by the Boltzmann equation, one needs to include additional terms into the classical equation such as removal events (e.g., chemical reactions), interactions with a background host medium, and presence of an external source \cite{bk:Spiga[1984],bk:BoffiSpiga[1982],bk:BoffiSpiga2[1982]}.

In \cite{bk:Spiga[1984]}, a particular solution of the spatially homogeneous Boltzmann equation with Maxwell molecules including removal was found. The authors of \cite{bk:SantosBrey[1985]} showed that using an equivalence transformation one can reduce the Boltzmann equation with removal to the classical Boltzmann equation.

In \cite{bk:Nonnenmacher[1984]}, using the method developed in \cite{bk:KrookWu[1976]}, a nonlinear second-order partial differential equation was derived from the Boltzmann equation for the spatially homogeneous case with a source function.  Applying the classical group analysis method, some forms of invariant solutions were found. However, because of the presence of a nonlocal term in the equation, the classical group analysis method cannot be applied to the equation. The results of \cite{bk:Nonnenmacher[1984]} were corrected in \cite{bk:GrigorievMeleshko[2012],bk:GrigorievMeleshkoSuriyawichitseranee[2014]}, where the authors obtained the determining equation by using the method developed for equations with nonlocal terms, and the complete group classification of the equation with a source was obtained.

The determining equation of the Fourier image of the spatially homogeneous and isotropic Boltzmann equation  with a source was studied in \cite{bk:GrigorievMeleshkoSuriyawichitseranee[2015],bk:SuriyawichitseraneeGrigorievMeleshko[2015],bk:LongAdisakAmornratGrigorievMeleshko[2017]}. The equation with $q=q(x,t)$ and $q=q(x,t,\varphi)$ was analyzed in \cite{bk:GrigorievMeleshkoSuriyawichitseranee[2015]} and \cite{bk:LongAdisakAmornratGrigorievMeleshko[2017]}, respectively, using the analysis which was applied in \cite{bk:GrigorievMeleshko[2012],bk:GrigorievMeleshkoSuriyawichitseranee[2014]}. Later, in \cite{bk:SuriyawichitseraneeGrigorievMeleshko[2015]}, it was shown that the group classification obtained in \cite{bk:GrigorievMeleshkoSuriyawichitseranee[2015]} is complete.

This paper is devoted to application of the method of preliminary group classification \cite{bk:LongAdisakAmornratGrigorievMeleshko[2017]} for group classification of the Boltzmann equation
\begin{equation}\label{eq:fullBoltzannEqWithSources}
	\frac{\partial f}{\partial t} + \textbf{v}\cdot \nabla_{\textbf{x}}f - J(f,f) = q
\end{equation}
with respect to the source term using the Lie group $L_{11}$. The function $q$ depends on both, the independent and dependent variables, i.e.,  $q=q(t,\textbf{x},\textbf{v},f)$.

\subsection{Preliminary group classification}

Besides complete group classification there is a method proposed in
\cite{bk:AkhatovGazizovIbragimov[1991]} which is called \textit{preliminary group classification}. Further development of this method for differential equations is given
in \cite{bk:IbragimovTorrisiValenti[1991],bk:BihloBihloPopovych[2011]}. The main idea of preliminary group classification is based on the study of only those extensions of the kernel of admitted Lie groups that are induced by the transformations from the corresponding
equivalence Lie group. The problem of finding inequivalent cases of
such extensions of symmetry then reduces to the classification of
inequivalent subgroups of the equivalence Lie group. In particular,
if a Lie group is finite-parameter, then one can use an optimal systems
of its subgroups.

\subsection{Structure of the paper}

The paper is separated into three parts. The first part of the paper is devoted to the general study of deriving determining equation of the Lie group admitted by nonlocal equation $\Phi(f)=q$ using the group properties of the homogeneous equation $\Phi(f)=0$: a strategy for constructing the determining equation for the source function $q$ is derived.

The developed strategy is applied to Eq. (\ref{eq:fullBoltzannEqWithSources}) in the next part. The determining equation for the function $q(x,y,z,u,v,w,t,f)$ for each subalgebra of the optimal system of subalgebras of the Lie algebra $L_{11}$ were studied. Obtaining the source function is illustrated by some examples. The complete results of the preliminary group classification are presented in Appendix \ref{AppendixGroupClassification}. It should be noted that $L_{11}$ is admitted by Eq. (\ref{eq:fullBoltzannEqWithSources}) only with the source term $q=Cf^{2}$, where $C$ is constant.

The third part of the paper provides representations of invariant solutions of Eq. (\ref{eq:fullBoltzannEqWithSources}). Complete results of the representations of invariant solutions are given in Appendix \ref{AppendixRepInvSol}. The reduced equations are also considered in this part.

\section{Method applied in the paper}\label{section:PreliminaryGroupClassificationOfTheBE&S}

The basic idea of the method is as follows. Consider an equation with nonlocal operator
\begin{equation}\label{IDEgeneral}
	\Phi(\textbf{x},f)=0,
\end{equation}
where $\textbf{x}=(x_{1},\dots,x_{n})$ is the vector of the independent variables, $f$ is the dependent variable, and $\Phi$ is a nonlocal operator acting on $f$. There are many situations where after group analysis of Eq. (\ref{IDEgeneral}), it is of interest to classify an inhomogeneous equation of the form
\begin{equation*}
	\Phi(\textbf{x},f)=q,
\end{equation*}
where $q$ is an arbitrary function\footnote{In many equations of mathematical physics the function $q$ plays the role of a source.} with respect to which a group classification required. The basic idea is to
exploit group properties of the homogeneous equation (\ref{IDEgeneral})
such as admitted Lie algebra, an optimal system of subalgebras of the Lie algebra and representations of invariant solutions.

Let
\begin{align}
	X=\xi^{i}(\textbf{x},f)\partial_{x_{i}}+\eta(\textbf{x},f)\partial_{f}
\end{align}
be a generator of a Lie group $G^{1}$:
\begin{align}\label{groupOfTrs}
	\bar{\textbf{x}}=\textbf{g}(\textbf{x},f;a),~\bar{f}=\psi(\textbf{x},f;a),
\end{align}
admitted by Eq. (\ref{IDEgeneral}), where $a$ is a parameter of the Lie group.

The property of $G^{1}$ to be admitted provides that
\begin{equation}\label{property:AdmittedIDEgeneral}
	\tilde{X}(\Phi)+\xi^{i} D_{i}(\Phi)=h_{X}\Phi, 
\end{equation}
where $D_{i}$ are the total derivatives with respect to $x_{i},~(i=1,2\dots,n)$, and  $\tilde{X}$ is the prolongation of the canonical Lie-B\"acklund operator which is equivalent to the generator $X$:
\begin{align}
	\tilde{X}=\tilde{\eta}\partial_{f}+D_{k}(\tilde{\eta})\partial_{f_{x_{k}}}+\cdots.
\end{align}
In general, $h_{X}$ is a functional, whereas for partial differential equations $h_{X}$ is a function of the independent and dependent variables and their derivatives. Here $\tilde{\eta}=\eta-\xi^{i}f_{x_{i}}$, and the actions of the derivatives $\partial_{f}$ and $\partial_{f_{x_{k}}}$ are considered in terms of the Fr\'echet derivatives.

For the preliminary group classification of the equation
\begin{equation}\label{IDE&Sgeneral}
	\Phi(\textbf{x},f)=q
\end{equation}
one uses the Lie algebra admitted by the homogeneous equation (\ref{IDEgeneral}). Here $q$ is an arbitrary function of the independent and dependent variables, i.e., $q=q(\textbf{x},f)$.

First, one defines the determining equation for a Lie algebra admitted by Eq. (\ref{IDE&Sgeneral}), and then this determining equation is simplified by using the property that the admitted generator belongs to a Lie algebra admitted by Eq. (\ref{IDEgeneral}).

According to the definition for a Lie group $G^{1}$ to be admitted \cite{bk:GrigMel[1986],bk:GrigorievIbragimovKovalevMeleshko2010}, one has the determining equation
\begin{equation}\label{determiningEq5}
	(\tilde{X}(\Phi-q))_{|(\ref{IDE&Sgeneral})}=0,
\end{equation}
which is further used for the group classification. Here $|(\ref{IDE&Sgeneral})$ means that the determining equation (\ref{determiningEq5}) is satisfied for any solutions of Eq. (\ref{IDE&Sgeneral}).

For the preliminary group classification studied in this paper the generator $X$ is assumed to belong to a Lie algebra admitted by the homogeneous equation. Hence, the generator $X$ satisfies the property (\ref{property:AdmittedIDEgeneral}). Using this property, one obtains that
\begin{align}
	\tilde{X}(\Phi-q)=\tilde{X}(\Phi)-\tilde{X}(q).
\end{align}
As the function $q$ only depends on the dependent and independent variables, then $\tilde{X}(q)=X(q)-\xi^{i}D_{i}(q)$. Hence, by virtue of (\ref{property:AdmittedIDEgeneral}),
\begin{align}\label{appliedBE&S}
	\tilde{X}(\Phi-q)=h_{X}\Phi-X(q)-\xi^{i}D_{i}(\Phi-q).
\end{align}
Considering Eq. (\ref{appliedBE&S}) on a solution of Eq. (\ref{IDE&Sgeneral}), one derives the determining equation,
\begin{equation}
	\tilde{h}_{X}q-X(q)=0,
\end{equation}\label{determiningEqGeneralCase}
where $\tilde{h}_{X}=(h_{X})_{|(\ref{IDE&Sgeneral})}$.

Thus, one comes to the following algorithm for a preliminary group classification of non-homogeneous equation (\ref{IDE&Sgeneral}) with the help of a Lie group admitted by the homogeneous equation (\ref{IDEgeneral}).

Let $L_{n}$ be a finite-dimensional Lie algebra and its basis generators $X_{1},X_{2},\dots,X_{n}$.
\begin{description}
	\item[Step 1] Take a subalgebra $L_{k},~(k\leq n)$ from an optimal system of subalgebras of $L_{n}$, with the basis generators $Y_{i}=c_{i}^{j}X_{j},~(i=1,2,\dots,k)$.
	\item[Step 2] Compute the prolongation $\tilde{Y}_{i}$ of the canonical Lie-B\"acklund operator corresponding to the generator $Y_{i}~(i=1,\dots,k)$.
	\item[Step 3] Solve the overdetermined system of $k$ partial differential equations
	\begin{equation}\label{determiningEqFor1subalgebra}
		\tilde{h}_{Y_{i}}q-Y_{i}(q)=0,~(i=1,\dots,k),
	\end{equation}
	where $\tilde{h}_{Y_{i}}=(h_{Y_{i}})_{|(\ref{IDE&Sgeneral})}$.
	The found solution provides a function $q(\textbf{x},f)$ such that Eq. (\ref{IDE&Sgeneral}) admits the Lie algebra $L_{k}$.
\end{description}

As the Lie algebra $L_{n}$ is determined by its basis generators $X_{1},\dots,X_{n}$, the functions $h_{Y_{i}}$ in Eq. (\ref{determiningEqFor1subalgebra}) can be also written through the functions $h_{X_{j}},~(j=1,\dots,n)$. In fact, because of the linearity of the Lie-B\"acklund operator,
\begin{align}\label{linearityOfOperator}
	\tilde{Y}_{i}&=c_{i}^{j}\tilde{X}_{j},
\end{align}
hence,
\begin{align}\label{determiningEq2}
	h_{Y_{i}}\Phi&=\tilde{Y}_{i}(\Phi)+\xi_{Y_{i}}^{k}D_{k}(\Phi)\nonumber\\
	&=c_{i}^{j}\tilde{X}_{j}(\Phi)+c_{i}^{j}\xi_{X_{j}}^{k}D_{k}(\Phi)\nonumber\\
	&=c_{i}^{j}(\tilde{X}_{j}(\Phi)+\xi_{X_{j}}^{k}D_{k}(\Phi))\nonumber\\
	&=c_{i}^{j}h_{X_{j}}\Phi.
\end{align}
Thus, the determining equation (\ref{determiningEqFor1subalgebra}) becomes
\begin{equation}\label{determiningEq3}
	c_{i}^{j}(\tilde{h}_{X_{j}}q-X_{j}(q))=0,~(i=1,\dots,k).
\end{equation}

\section{Group classification of source functions}

This section gives examples which illustrate the application of the above strategy. Complete results of the preliminary group classification are presented in Appendix \ref{AppendixGroupClassification}.

Let $\Phi$ be the operator
\begin{equation}
	\Phi(f)=f_{t}+uf_{x}+vf_{y}+wf_{z}-J(f,f),
\end{equation}
where $J(f,f)$ is the collision integral (\ref{collisionIntegral}).

As mentioned in the Introduction, the Boltzmann equation $\Phi(f)=0$ admits the Lie algebra $L_{11}$ with the basis generators (\ref{basisGeneratorsOfL11}). The Lie algebra $L_{11}$ is isomorphic to the Lie algebra $L_{11}^{g}$ admitted by the gas dynamics equations \cite{bk:GrigMel[1999],bk:GrigorievMeleshko[2001]}. In \cite{bk:Ovsiannikov[1994b]}, an optimal system of subalgebras of $L_{11}^{g}$ is constructed. Because of the isomorphism of $L_{11}$ and $L_{11}^{g}$, one can apply the method described above for solving the preliminary group classification problem of the Boltzmann equation with a source,
\begin{align}\label{eq:BoltzmannWithSource}
	\Phi(f)=q.
\end{align}

First, one provides definitions, equations, and their variables which are useful for the next section.

Let $f=f_{0}(x,y,z,u,v,w,t)$ be a function; the transformed function is denoted as $\bar{f}=f_{a}(\bar{x},\bar{y},\bar{z},\bar{u},\bar{v},\bar{w},\bar{t})$.
The collision integral for the transformed function has the form
\begin{equation}\label{transformedCollisionTerm}
	J(\bar{f},\bar{f})=\int\limits_{\mathbb{R}^{3}}\int\limits_{\mathbf{S}^{2}}B(\bar{g},\theta_{1})(\bar{f}^{*}\bar{f}_{1}^{*}-\bar{f}\bar{f}_{1})d\textbf{n}d\textbf{w}
\end{equation}
where
\begin{align}\label{transformedVariablesInCollision}
	&\bar{f}=\bar{f}(\bar{\textbf{x}},\bar{\textbf{v}},\bar{t}),~\bar{f}_{1}=\bar{f}(\bar{\textbf{x}},\textbf{w},\bar{t}),~\bar{f}^{*}=\bar{f}(\bar{\textbf{x}},\bar{\textbf{v}}^{*},\bar{t}),~\bar{f}_{1}^{*}=\bar{f}(\bar{\textbf{x}},\bar{\textbf{w}}^{*},\bar{t}),\nonumber\\
	&\bar{\textbf{x}}=(\bar{x},\bar{y},\bar{z}),~\bar{\textbf{v}}=(\bar{u},\bar{v},\bar{w}),~\textbf{w}=(u_{1},v_{1},w_{1}),~\bar{\textbf{v}}^{*}=\frac{1}{2}(\bar{\textbf{v}}+\textbf{w}+\bar{g}\textbf{n}),\nonumber\\
	&\bar{\textbf{w}}^{*}=\frac{1}{2}(\bar{\textbf{v}}+\textbf{w}-\bar{g}\textbf{n}),~\bar{\textbf{g}}=\bar{\textbf{v}}-\textbf{w},~\bar{g}=\|\bar{\textbf{g}}\|_{2}.
\end{align}

In this section one shows that Eq. (\ref{determiningEq3}) for the Boltzmann equation (\ref{eq:BoltzmannWithSource}) is
\begin{equation}\label{determiningEq4}
	c_{i}^{11}(2q)+c_{i}^{j}(X_{j}(q))=0,~(i=1,\dots,k).
\end{equation}

\subsection{The functions $\tilde{h}_{X_{j}},~(j=1,2,\dots,11)$}

First, it should be noted that for deriving a function transformed by the Lie group corresponding to (\ref{basisGeneratorsOfL11}) there is no necessity to apply the inverse function theorem. In this case for finding the functions $h_{X_{j}}$ one can use the algorithm which is illustrated here by the generator $X_{11}$.

The group of transformations corresponding to the generator $X_{11}$ is
\begin{equation}\label{TransformationGroup11}
	\bar{x}=xe^{a},~\bar{y}=ye^{a},~\bar{z}=ze^{a},~\bar{u}=u,~\bar{v}=v,~\bar{w}=w,~\bar{t}=te^{a},~\bar{f}=fe^{-a}.
\end{equation}
This group of transformations maps a function $f=f_{0}(x,y,z,u,v,w,t)$ to the function $\bar{f}=f_{a}(\bar{x},\bar{y},\bar{z},\bar{u},\bar{v},\bar{w},\bar{t})$, where the transformed function is determined by the formula
\begin{equation}\label{fNewAndfRelation11}
	\bar{f}(\bar{x},\bar{y},\bar{z},\bar{u},\bar{v},\bar{w},\bar{t})=e^{-a}f_{0}(\bar{x}e^{-a},\bar{y}e^{-a},\bar{z}e^{-a},\bar{u},\bar{v},\bar{w},\bar{t}e^{-a}).
\end{equation}
It follows that
\begin{align}
	(\Phi(\bar{f}))(\bar{x},\bar{y},\bar{z},\bar{u},\bar{v},\bar{w},\bar{t})&=e^{-2a}(f_{0t}+uf_{0x}+vf_{0y}+wf_{0z}\nonumber\\
	&-J(f_{0},f_{0}))(\bar{x}e^{-a},\bar{y}e^{-a},\bar{z}e^{-a},\bar{u},\bar{v},\bar{w},\bar{t}e^{-a})\nonumber\\
	&=e^{-2a}(\Phi(f_{0}))(\bar{x}e^{-a},\bar{y}e^{-a},\bar{z}e^{-a},\bar{u},\bar{v},\bar{w},\bar{t}e^{-a}).
\end{align}
Differentiating the latter relation with respect to $a$ and setting $a=0$, one derives
\begin{align}
	\tilde{X}_{11}(\Phi)&=-2\Phi-xD_{x}(\Phi)-yD_{y}(\Phi)-zD_{z}(\Phi)-tD_{t}(\Phi).
\end{align}
As
\begin{align}
	h_{X_{11}}\Phi&=\tilde{X}_{11}(\Phi)+(xD_{x}(\Phi)+yD_{y}(\Phi)+zD_{z}(\Phi)+tD_{t}(\Phi))=-2\Phi,
\end{align}
one finds that
\begin{equation}\label{hX11onManifold}
	\tilde{h}_{X_{11}}=-2.
\end{equation}
Similarly, one obtains that
\begin{equation}
	\tilde{h}_{X_{j}}=0,~(j=1,2,\dots,10).
\end{equation}

\subsection{Illustrative examples of finding a source function}\label{obtainingSourceFunc}

For solving differential equations corresponding to a subalgebra of the Lie algebra $L_{11}$ one needs to find their integrals. Among these subalgebras there are subalgebras containing the generators of the rotations $X_{7}$, $X_{8}$, and $X_{9}$. In some of these cases it is appropriate to write the determining equations (\ref{determiningEq4}) in the cylindrical or spherical coordinate systems. It should also be mentioned that the necessity of the changing the coordinate system depends on the complexity of the system of determining equations. As different coordinate systems are used, we provide illustrative examples in all of these coordinate systems. The changes of variables in the cylindrical and spherical coordinate systems are provided in Appendix \ref{AppendixChangeofVarInCylSph}.

\subsubsection{A source function corresponding to the subalgebra 1.8 : $\{X_{11}\}$}
Substituting
\begin{align*}
	c_{1}^{11}=1,~c_{1}^{j}=0,~j\neq11
\end{align*}
into (\ref{determiningEq4}), one obtains the partial differential equation
\begin{align}\label{ex1.8_eqQ1}
	2q-fq_{f}+tq_{t}+xq_{x}+yq_{y}+zq_{z}=0.
\end{align}
The general solution of Eq. (\ref{ex1.8_eqQ1}) has the form
\begin{equation*}
	q=t^{-2}\Psi(\frac{x}{t},\frac{y}{t},\frac{z}{t},u,v,w,ft),
\end{equation*}
where $\Psi$ is an arbitrary function of 7 variables.

\subsubsection{A source function corresponding to the subalgebra 1.2 : $\{\beta X_{4}+X_{7}\},~\beta\neq0$}
Substituting
\begin{align*}
	c_{1}^{4}=\beta,~c_{1}^{7}=1,~c_{1}^{j}=0,~j\neq4,7
\end{align*}
into (\ref{determiningEq4}), one obtains the equation
\begin{align}\label{ex1.2_eqQ}
	\beta q_{u}+\beta tq_{x}-wq_{v}+vq_{w}-zq_{y}+yq_{z}=0.
\end{align}
Using change of variables in cylindrical coordinates, Eq. (\ref{ex1.2_eqQ}) becomes
\begin{align*}
	\beta q_{u}+\beta tq_{x}+q_{\theta}=0.
\end{align*}
The general solution of Eq. (\ref{ex1.2_eqQ}) in the cylindrical coordinate system has the form
\begin{equation*}
	q=\Psi(t,r,\beta\theta-\frac{x}{t},u-\frac{x}{t},V,W,f),
\end{equation*}
where $\Psi$ is an arbitrary function of 7 variables.

\subsubsection{A source function corresponding to the subalgebra 3.8 : $\{X_{7},X_{8},X_{9}\}$}\label{exSourceFncForSub3.8}
Substituting the coefficients
\begin{align*}
	&c_{1}^{7}=1,~c_{1}^{j}=0,~j\neq7;~c_{2}^{8}=1,~c_{2}^{j}=0,~j\neq8;~c_{3}^{9}=1,~c_{3}^{j}=0,~j\neq9
\end{align*}
into (\ref{determiningEq4}), one obtains the system of three partial differential equations:
\begin{eqnarray}
	-wq_{v}+vq_{w}-zq_{y}+yq_{z}=0,\label{ex3.8_eqQ1}\\
	wq_{u}-uq_{w}+zq_{x}-xq_{z}=0,\label{ex3.8_eqQ2}\\
	-vq_{u}+uq_{v}-yq_{x}+xq_{y}=0.\label{ex3.8_eqQ3}
\end{eqnarray}
Using a change of variables to the spherical coordinate system, Eqs. (\ref{ex3.8_eqQ1})-(\ref{ex3.8_eqQ3}) become
\begin{eqnarray}
	-\sin(\varphi)q_{\theta}-\cos(\varphi)\cot(\theta)q_{\varphi}-\frac{\cos(\varphi)}{\sin(\theta)}(Wq_{V}-Vq_{W})=0,\label{ex3.8_eqQ1New}\\
	\cos(\varphi)q_{\theta}-\sin(\varphi)\cot(\theta)q_{\varphi}-\frac{\sin(\varphi)}{\sin(\theta)}(Wq_{V}-Vq_{W})=0,\label{ex3.8_eqQ2New}\\
	q_{\varphi}=0.\label{ex3.8_eqQ3New}
\end{eqnarray}
Because of Eq. (\ref{ex3.8_eqQ3New}), $q$ does not depend on $\varphi$, i.e., $q=q(r,\theta,U,V,W,t,f)$. Eqs. (\ref{ex3.8_eqQ1New})-(\ref{ex3.8_eqQ2New}) become
\begin{align*}
	&\sin(\theta)\sin(\varphi)q_{\theta}+\cos(\varphi)(Wq_{V}-Vq_{W})=0,\\
	&\sin(\theta)\cos(\varphi)q_{\theta}-\sin(\varphi)(Wq_{V}-Vq_{W})=0.
\end{align*}
The latter system of equations can be rewritten in the form
\begin{align*}
	&q_{\theta}=0,\\
	&Wq_{V}-Vq_{W}=0.
\end{align*}
The general solution of Eqs. (\ref{ex3.8_eqQ1})-(\ref{ex3.8_eqQ3}) in the spherical coordinate system has the form
\begin{equation*}
	q=\Psi(t,r,U,\sqrt{V^{2}+W^{2}},f),
\end{equation*}
where $\Psi$ is an arbitrary function of 5 variables.

\section{Representations of invariant solutions and the reduced Boltzmann equation}

This section is devoted to obtaining representations of invariant solutions of the Boltzmann equation with a source function. Besides discussing representations of invariant solutions, this section is also devoted to finding reduced equations. It should be noted that for some of representations of invariant solutions, the problem of obtaining a reduced equation is extremely difficult.

In the previous section source functions $q_{k}$ are obtained for all subalgebras $L_{k}$ of the optimal system of the Lie algebra $L_{11}$ such that the Boltzmann equation with the source function $q_{k}$ admits the Lie algebra $L_{k}$.

Let $L$ be any subalgebra of $L_{k}$. As $L_{k}$ is a subalgebra of $L_{11}$, then $L$ is a subalgebra of $L_{11}$. According to the definition of an optimal system of subalgebras, $L$ is equivalent to one of subalgebras of the optimal system of subalgebras of the Lie algebra $L_{11}$, say $\tilde{L}$. Hence, invariant solutions with respect to these subalgebras $L$  and $\tilde{L}$ are equivalent.

As the set of representations of invariant solutions of all subalgebras from the optimal system of subalgebras will be found, and because solutions invariant with respect to the subalgebra $L$ and the subalgebra $\tilde{L}$ are equivalent, then the representations of all invariant solutions will be given.

In this section representations of invariant solutions will be constructed for all subalgebras of the optimal system. This study will provide representations of all possible invariant solutions of the Boltzmann equation of the form (\ref{eq:BoltzmannWithSource}).

\subsection{Illustrative examples of obtaining the representation of invariant solution}\label{obtainingRepInvarSol}
In this section examples which illustrate the method of finding a representation of invariant solutions in the Cartesian, cylindrical, and spherical coordinate systems are given.

\subsubsection{A representation of invariant solutions corresponding to the subalgebra 1.8 : $\{X_{11}\}$}
For this subalgebra, for finding invariants one needs to solve the equation
\begin{align*}
	X_{11}(J)=0,
\end{align*}
i.e.,
\begin{align}\label{ex1.8_eqJ}
	tJ_{t}+xJ_{x}+yJ_{y}+zJ_{z}-fJ_{f} = 0.
\end{align}
The independent invariants of Eq. (\ref{ex1.8_eqJ}) are
\begin{align*}
	J_{1}=\frac{x}{t},J_{2}=\frac{y}{t},J_{3}=\frac{z}{t},J_{4}=u,J_{5}=v,J_{6}=w,J_{7}=ft.
\end{align*}
A representation of invariant solutions for this subalgebra has the form
\begin{equation*}\label{repInvariantSol1.8}
	f=t^{-1}\Omega(\frac{x}{t},\frac{y}{t},\frac{z}{t},u,v,w),
\end{equation*}
where $\Omega$ is an arbitrary function of 6 variables.

\subsubsection{A representation of invariant solutions corresponding to the subalgebra 1.2 : $\{\beta X_{4}+X_{7}\},~\beta\neq0$}
In the cylindrical coordinate system, the generator of this subalgebra is
\begin{equation*}
	\beta X_{4c}+X_{7c},
\end{equation*}
where $X_{4c}=t\partial_{x}+\partial_{u},~X_{7c}=\partial_{\theta}$. For this subalgebra, for finding invariants one needs to solve the equation
\begin{align}\label{ex1.2_eqJNew}
	\beta(tJ_{x}+J_{u})+J_{\theta}=0,
\end{align}
where $J=J(x,r,\theta,u,V,W,t,f)$.
The independent invariants of Eq. (\ref{ex1.2_eqJNew}) are
\begin{align*}
	J_{1}=t,J_{2}=r,J_{3}=\beta\theta-\frac{x}{t},J_{4}=u-\frac{x}{t},J_{5}=V,J_{6}=W,J_{7}=f,
\end{align*}
and a representation of invariant solutions in the cylindrical coordinate system for this subalgebra is
\begin{equation*}\label{repInvariantSol1.2}
	f=\Omega(t,r,\beta\theta-\frac{x}{t},u-\frac{x}{t},V,W),
\end{equation*}
where $\Omega$ is an arbitrary function of 6 independent variables.

\subsubsection{A representation of invariant solutions corresponding to the subalgebra 3.8 : $\{X_{7},X_{8},X_{9}\}$}
In the spherical coordinate system, the generators of this subalgebra are
\begin{align*}
	X_{7s}&=-\sin(\varphi)\partial_{\theta}-\cos(\varphi)\cot(\theta)\partial_{\varphi}-\frac{\cos(\varphi)}{\sin(\theta)}(W\partial_{V}-V\partial_{W}),\\
	X_{8s}&=\cos(\varphi)\partial_{\theta}-\sin(\varphi)\cot(\theta)\partial_{\varphi}-\frac{\sin(\varphi)}{\sin(\theta)}(W\partial_{V}-V\partial_{W}),\\
	X_{9s}&=\partial_{\varphi}.
\end{align*}
For this subalgebra, one needs to solve the system of equations
\begin{eqnarray}
	-\sin(\varphi)J_{\theta}-\cos(\varphi)\cot(\theta)J_{\varphi}-\frac{\cos(\varphi)}{\sin(\theta)}(WJ_{V}-VJ_{W})=0,\label{ex3.8_eqJ1New}\\
	\cos(\varphi)J_{\theta}-\sin(\varphi)\cot(\theta)J_{\varphi}-\frac{\sin(\varphi)}{\sin(\theta)}(WJ_{V}-VJ_{W})=0,\label{ex3.8_eqJ2New}\\
	J_{\varphi}=0,\label{ex3.8_eqJ3New}
\end{eqnarray}
where $J=J(r,\theta,\varphi,U,V,W,t,f)$.
The independent invariants of Eqs. (\ref{ex3.8_eqJ1New})-(\ref{ex3.8_eqJ3New}) are
\begin{align*}
	J_{1}=t,J_{2}=r,J_{3}=U,J_{4}=\sqrt{V^{2}+W^{2}},J_{5}=f.
\end{align*}
Thus, a representation of invariant solutions in the spherical coordinate system for this subalgebra is
\begin{equation*}
	f=\Omega(t,r,U,\sqrt{V^{2}+W^{2}}),
\end{equation*}
where $\Omega$ is an arbitrary function of 4 variables.

\subsection{Illustrative examples of obtaining reduced equations}

The full Boltzmann equation with a source function in cylindrical and spherical coordinate systems are provided in Appendix \ref{AppendixChangeofVarInCylSph}. Substituting the source function and the representation of an invariant solution found above into the Boltzmann equation, one obtains the reduced Boltzmann equation with a source function. In this section some examples of obtaining a reduced equation with a source function in Cartesian, and cylindrical coordinate systems are presented.

\subsubsection{Reduced equation for the subalgebra 1.8 : $\{X_{11}\}$}
A source function and a representation of invariant solutions corresponding to subalgebra  $\{X_{11}\}$ are
\begin{align*}
	q=t^{-2}\Psi(p_{1},p_{2},p_{3},u,v,w,ft),~f=t^{-1}\Omega(p_{1},p_{2},p_{3},u,v,w),
\end{align*}
where $p_{1}=\frac{x}{t},~p_{2}=\frac{y}{t},~p_{3}=\frac{z}{t}$.
The differential part of the Boltzmann equation (\ref{eq:fullBoltzannEqWithSources}) becomes
\begin{equation}\label{ex1.8differentialPart}
	f_{t}+uf_{x}+vf_{y}+wf_{z}=t^{-2}(-\Omega+(u-p_{1})\Omega_{p_{1}}+(v-p_{2})\Omega_{p_{2}}+(w-p_{3})\Omega_{p_{3}}).
\end{equation}
As $f=t^{-1}\Omega(\frac{x}{t},\frac{y}{t},\frac{z}{t},u,v,w)$, then
\begin{align*}
	f_{1}&=t^{-1}\Omega(\frac{x}{t},\frac{y}{t},\frac{z}{t},u_{1},v_{1},w_{1}),\nonumber\\
	f^{*}&=t^{-1}\Omega(\frac{x}{t},\frac{y}{t},\frac{z}{t},\frac{1}{2}(u+u_{1}+gn_{1}),\frac{1}{2}(v+v_{1}+gn_{2}),\frac{1}{2}(w+w_{1}+gn_{3})),\nonumber\\
	f_{1}^{*}&=t^{-1}\Omega(\frac{x}{t},\frac{y}{t},\frac{z}{t},\frac{1}{2}(u+u_{1}-gn_{1}),\frac{1}{2}(v+v_{1}-gn_{2}),\frac{1}{2}(w+w_{1}-gn_{3})),
\end{align*}
and
\begin{align}\label{sourceFnc1.8RecalledNew}
	q=t^{-2}\Psi(p_{1},p_{2},p_{3},u,v,w,\Omega).
\end{align}
The collision term in (\ref{eq:fullBoltzannEqWithSources}) becomes
\begin{equation}\label{ex1.8CollisionTerm}
	J(f,f)=t^{-2}\int\limits_{\mathbb{R}^{3}}\int\limits_{\mathbf{S}^{2}}B(g,\theta_{1})(\Omega^{*}\Omega_{1}^{*}-\Omega\Omega_{1})\sin{\theta_{1}}\,d\theta_{1}\,d\epsilon\,du_{1}\,dv_{1}\,dw_{1},
\end{equation}
where
\begin{align*}
	\Omega&=\Omega(p_{1},p_{2},p_{3},u,v,w),~\Omega_{1}=\Omega(p_{1},p_{2},p_{3},u_{1},v_{1},w_{1}),\nonumber\\
	\Omega^{*}&=\Omega(p_{1},p_{2},p_{3},\frac{1}{2}(u+u_{1}+gn_{1}),\frac{1}{2}(v+v_{1}+gn_{2}),\frac{1}{2}(w+w_{1}+gn_{3})),\nonumber\\
	\Omega_{1}^{*}&=\Omega(p_{1},p_{2},p_{3},\frac{1}{2}(u+u_{1}-gn_{1}),\frac{1}{2}(v+v_{1}-gn_{2}),\frac{1}{2}(w+w_{1}-gn_{3}))\nonumber\\
	g&=\sqrt{(u-u_{1})^{2}+(v-v_{1})^{2}+(w-w_{1})^{2}},~p_{1}=\frac{x}{t},~p_{2}=\frac{y}{t},~p_{3}=\frac{z}{t}.
\end{align*}
Substituting (\ref{ex1.8differentialPart}),~(\ref{sourceFnc1.8RecalledNew}), and (\ref{ex1.8CollisionTerm})  into the Boltzmann equation (\ref{eq:fullBoltzannEqWithSources}), and multiplying the obtained equation by $t^{2}$, one derives the reduced equation
\begin{align*}
	-\Omega&+(u-p_{1})\Omega_{p_{1}}+(v-p_{2})\Omega_{p_{2}}+(w-p_{3}\Omega_{p_{3}})-J(\Omega,\Omega)=\Psi(p_{1},p_{2},p_{3},u,v,w,\Omega),
\end{align*}
where
\begin{align*}
	J(\Omega,\Omega)=\int\limits_{\mathbb{R}^{3}}\int\limits_{\mathbf{S}^{2}}B(g,\theta_{1})(\Omega^{*}\Omega_{1}^{*}-\Omega\Omega_{1})\sin{\theta_{1}}\,d\theta_{1}\,d\epsilon\,du_{1}\,dv_{1}\,dw_{1}.
\end{align*}

\subsubsection{Reduced equation for the subalgebra 1.2 : $\{\beta X_{4}+X_{7}\},~\beta\neq0$}
For this case a source function and a representation of invariant solutions are
\begin{align*}
	q=\Psi(t,r,p,\tilde{u},V,W,f),~f=\Omega(t,r,p,\tilde{u},V,W),
\end{align*}
where $p=\beta\theta-\frac{x}{r},~\tilde{u}=u-\frac{x}{t}$. The differential part of the Boltzmann equation (\ref{eq:fullBoltzannEqInCylindrical}) becomes
\begin{align}\label{ex1.2differentialPart}
	f_{t}+uf_{x}+Vf_{r}+\frac{W}{r}f_{\theta}+\frac{W^{2}}{r}f_{V}-\frac{WV}{r}f_{W}&=\Omega_{t}+(\frac{\beta W}{r}-\frac{\tilde{u}}{t})\Omega_{p}+V\Omega_{r}-\frac{\tilde{u}}{t}\Omega_{\tilde{u}}\nonumber\\
	&+\frac{W^{2}}{r}\Omega_{V}-\frac{VW}{r}\Omega_{W}.
\end{align}
As $f=\Omega(t,r,\beta\theta-\frac{x}{r},u-\frac{x}{t},V,W)$, then
\begin{align}\label{ex1.2fInCollisionTerm}
	f_{1}&=\Omega(t,r,\beta\theta-\frac{x}{r},u_{1}-\frac{x}{t},V_{1},W_{1}),\nonumber\\
	f^{*}&=\Omega(t,r,\beta\theta-\frac{x}{r},\frac{1}{2}((u-\frac{x}{t})+(u_{1}-\frac{x}{t})+g_{c}n_{1c}),\frac{1}{2}(V+V_{1}+g_{c}n_{2c}),\frac{1}{2}(W+W_{1}+g_{c}n_{3c})),\nonumber\\
	f_{1}^{*}&=\Omega(t,r,\beta\theta-\frac{x}{r},\frac{1}{2}((u-\frac{x}{t})+(u_{1}-\frac{x}{t})-g_{c}n_{1c}),\frac{1}{2}(V+V_{1}-g_{c}n_{2c}),\nonumber\\
	&\frac{1}{2}(W+W_{1}-g_{c}n_{3c})),
\end{align}
and
\begin{align}\label{sourceFnc1.2RecalledNew}
	q=\Psi(t,r,p,\tilde{u},V,W,\Omega).
\end{align}
Substituting (\ref{ex1.2fInCollisionTerm}) into the collision integral of the Boltzmann equation (\ref{eq:fullBoltzannEqInCylindrical}), and using the change of variable: $\tilde{u}_{1}=u_{1}-\frac{x}{t}$, one obtains
\begin{align}\label{ex1.2CollisionTerm}
	J(f,f)=\int\limits_{\mathbb{R}^{3}}\int\limits_{\mathbf{S}^{2}}B(g_{c},\theta_{1})(\Omega^{*}\Omega_{1}^{*}-\Omega\Omega_{1})\sin{\theta_{1}}\,d\theta_{1}\,d\epsilon\,d\tilde{u}_{1}\,dV_{1}\,dW_{1},
\end{align}
where
\begin{align*}
	&\Omega=\Omega(t,r,p,\tilde{u},V,W),~\Omega_{1}=\Omega(t,r,p,\tilde{u}_{1},V_{1},W_{1}),\nonumber\\
	&\Omega^{*}=\Omega(t,r,p,\frac{1}{2}(\tilde{u}+\tilde{u}_{1}+g_{c}n_{1c}),\frac{1}{2}(V+V_{1}+g_{c}n_{2c}),\frac{1}{2}(W+W_{1}+g_{c}n_{3c})),\nonumber\\
	&\Omega_{1}^{*}=\Omega(t,r,p,\frac{1}{2}(\tilde{u}+\tilde{u}_{1}-g_{c}n_{1c}),\frac{1}{2}(V+V_{1}-g_{c}n_{2c}),\frac{1}{2}(W+W_{1}-g_{c}n_{3c})),\nonumber\\
	&g_{c}=\sqrt{(\tilde{u}-\tilde{u}_{1})^{2}+(V-V_{1})^{2}+(W-W_{1})^{2}},~p=\beta\theta-\frac{x}{r},~\tilde{u}=u-\frac{x}{t}.
\end{align*}
Substituting (\ref{ex1.2differentialPart}), (\ref{sourceFnc1.2RecalledNew}), and (\ref{ex1.2CollisionTerm}) into the Boltzmann equation in the cylindrical coordinate system, the reduced equation is
\begin{align*}
	\Omega_{t}+(\frac{\beta W}{r}-\frac{\tilde{u}}{t})\Omega_{p}+V\Omega_{r}-\frac{\tilde{u}}{t}\Omega_{\tilde{u}}+\frac{W^{2}}{r}\Omega_{V}&-\frac{VW}{r}\Omega_{W}-J(\Omega,\Omega)\nonumber\\
	&=\Psi(t,r,p,\tilde{u},V,W,\Omega),
\end{align*}
where
\begin{align*}
	J(\Omega,\Omega)=\int\limits_{\mathbb{R}^{3}}\int\limits_{\mathbf{S}^{2}}B(g_{c},\theta_{1})(\Omega^{*}\Omega_{1}^{*}-\Omega\Omega_{1})\sin{\theta_{1}}\,d\theta_{1}\,d\epsilon\,d\tilde{u}_{1}\,dV_{1}\,dW_{1}.
\end{align*}

\section{Conclusion}
The present paper gives a preliminary group classification of the full Boltzmann equation with a source term
\begin{equation}\label{ccsChpBE&S}
	f_{t}+uf_{x}+vf_{y}+wf_{z}-J(f,f)=q,
\end{equation}
where $q=q(x,y,z,u,v,w,t,f)$.
For the preliminary group classification the Lie algebra $L_{11}$ admitted by the classical Boltzmann equation without the a source term is applied. All source functions such that the Boltzmann equation (\ref{ccsChpBE&S}) admits a subalgebra of $L_{11}$ are obtained. Representations of all invariant solutions are given in the paper.

\ack
AK thanks Walailak University, and Royal Thai Government Scholarship through the Ministry of 
Science and Technology of Thailand for financial support.
SVM and YuNG acknowledge RFFI (code of the project 17-01-00209a) for a partial financial support.

\appendix

\section{Cylindrical coordinates}\label{AppendixChangeofVarInCylSph}
\indent
Consider the following change of variables \cite{bk:Ovsiannikov[1994b]}
\begin{align*}
	z=r\sin{\theta},~y=r\cos{\theta},~v=V\cos{\theta}-W\sin{\theta},~w=V\sin{\theta}+W\cos{\theta},
\end{align*}
where $\theta \in \left[0,2\pi\right),~r\geq 0$.

The basis generators in cylindrical coordinates are
\begin{align*}
	X_{1c}&=\partial_{x},~
	X_{2c}=\cos{\theta}\partial_{r}-\frac{\sin{\theta}}{r}\partial_{\theta}-\frac{W\sin{\theta}}{r}\partial_{V}+\frac{V\sin{\theta}}{r}\partial_{W},\nonumber\\
	X_{3c}&=\sin{\theta}\partial_{r}+\frac{\cos{\theta}}{r}\partial_{\theta}+\frac{W\cos{\theta}}{r}\partial _{V}-\frac{V\cos{\theta}}{r}\partial_{W},~
	X_{4c}=t \partial_{x} + \partial_{u},\nonumber\\
	X_{5c}&=t\cos{\theta}\partial_{r}-\frac{t\sin{\theta}}{r}\partial_{\theta}+(\cos{\theta}-\frac{tW\sin{\theta}}{r})\partial_{V}+(\frac{tV\sin{\theta}}{r}-\sin{\theta})\partial_{W},\nonumber\\
	X_{6c}&=t\sin{\theta}\partial_{r}+\frac{t\cos{\theta}}{r}\partial_{\theta}+(\sin{\theta}+\frac{tW\cos{\theta}}{r})\partial_{V}+(\cos{\theta}-\frac{tV\cos{\theta}}{r})\partial_{W},~
	X_{7c}=\partial_{\theta},\nonumber\\
	X_{8c}&=r\sin{\theta}\partial_{x}-x\sin{\theta}\partial_{r}-\frac{x\cos{\theta}}{r}\partial_{\theta}+(V\sin{\theta}+W\cos{\theta})\partial_{u}\nonumber\\
	&+(-u\sin{\theta}-\frac{xW\cos{\theta}}{r})\partial_{V}+(\frac{xV\cos{\theta}}{r}-u\cos{\theta})\partial_{W},\nonumber\\
	X_{9c}&=-r\cos{\theta}\partial_{x}+x\cos{\theta}\partial_{r}-\frac{x\sin{\theta}}{r}\partial_{\theta}+(-V\cos{\theta}+W\sin{\theta})\partial_{u}\nonumber\\
	&+(u\cos{\theta}-\frac{xW\sin{\theta}}{r})\partial_{V}-u\sin{\theta}\partial_{W},\nonumber\\
	X_{10c}&=\partial_{t},~
	X_{11c}=t\partial_{t}+x\partial_{x}+r\partial_{r}-f\partial_{f}.
\end{align*}

The Boltzmann equation in cylindrical coordinates is
\begin{equation*}
	f_{t}+uf_{x}+Vf_{r}+\frac{W}{r}f_{\theta}+\frac{W^{2}}{r}f_{V}-\frac{WV}{r}f_{W}=J(f,f),
\end{equation*}
where $f=f(x,r,\theta,u,V,W,t)$, and the collision integral
\begin{equation*}
	J(f,f)=\int\limits_{\mathbb{R}^{3}}\int\limits_{S^{2}}{B(g_{c},\theta_{1})(f^{*}f_{1}^{*}-ff_{1})\sin{\theta_{1}}}\,d\theta_{1}\,d\epsilon\,du_{1}\,dV_{1}\,dW_{1}.
\end{equation*}
Here
\begin{align*}
	&f=f(x,r,\theta,u,V,W,t),~f_{1}=f(x,r,\theta,u_{1},V_{1},W_{1},t),\nonumber\\
	&f^{*}=f(x,r,\theta,u^{*},V^{*},W^{*},t),~f_{1}^{*}=f(x,r,\theta,u_{1}^{*},V_{1}^{*},W_{1}^{*},t),
\end{align*}
and
\begin{align*}
	&u^{*}=\frac{1}{2}(u+u_{1}+g_{c}n_{1c}),~V^{*}=\frac{1}{2}(V+V_{1}+g_{c}n_{2c}),~W^{*}=\frac{1}{2}(W+W_{1}+g_{c}n_{3c}),\nonumber\\
	&u_{1}^{*}=\frac{1}{2}(u+u_{1}-g_{c}n_{1c}),~V_{1}^{*}=\frac{1}{2}(V+V_{1}-g_{c}n_{2c}),~W_{1}^{*}=\frac{1}{2}(W+W_{1}-g_{c}n_{3c}),\nonumber\\
	&\textbf{g}_{c}=((u-u_{1}),(V-V_{1}),(W-W_{1})),~g_{c}=\|\textbf{g}_{c}\|_{2},\\
	&\textbf{n}_{c}=(n_{1c},n_{2c},n_{3c})=(\cos{\theta_{1}},\cos{(\epsilon-\theta)}\sin{\theta_{1}},\sin{(\epsilon-\theta)}\sin{\theta_{1}}).
\end{align*}
\textit{Remark:}~$\textbf{n}:=(\cos{\theta_{1}},\sin{\theta_{1}}\cos{\epsilon},\sin{\theta_{1}}\sin{\epsilon})$,~$g_{c}=g$, and $\textbf{g}_{c}\cdot\textbf{n}_{c}=\textbf{g}\cdot\textbf{n}$.

\section{Spherical coordinates}

Consider the following change of variables \cite{bk:Ovsiannikov[1994b]}
\begin{align*}
	&x=r\sin{\theta}\cos{\varphi},y=r\sin{\theta}\sin{\varphi},z=r\cos{\theta},
	u=U\sin{\theta}\cos{\varphi}+V\cos{\theta}\cos{\varphi}-W\sin{\varphi},\nonumber\\
	&v=U\sin{\theta}\sin{\varphi}+V\cos{\theta}\sin{\varphi}+W\cos{\varphi},
	w=U\cos{\theta}-V\sin{\theta},
\end{align*}
where $\varphi \in \left[0,\pi\right),~\theta \in \left[0,2\pi\right)$.

Some basis generators in spherical coordinates are
\begin{align*}
	X_{7s}&=-\sin(\varphi)\partial_{\theta}-\cos(\varphi)\cot(\theta)\partial_{\varphi}-\frac{\cos(\varphi)}{\sin(\theta)}(W\partial_{V}-V\partial_{W}),\nonumber\\
	X_{8s}&=\cos(\varphi)\partial_{\theta}-\sin(\varphi)\cot(\theta)\partial_{\varphi}-\frac{\sin(\varphi)}{\sin(\theta)}(W\partial_{V}-V\partial_{W}),\nonumber\\
	X_{9s}&=\partial_{\varphi},~
	X_{10s}=\partial_{t},~
	X_{11s}=t\partial_{t}+r\partial_{r}-f\partial_{f}.
\end{align*}

The Boltzmann equation in spherical coordinates is
\begin{equation*}
	f_{t}+Uf_{r}+\frac{W}{r\sin{\theta}}f_{\varphi}+\frac{V}{r}f_{\theta}+\frac{V^{2}+W^{2}}{r}f_{U}+\frac{W^{2}\cot{\theta}-UV}{r}f_{V}-\frac{W(U+V\cot{\theta})}{r}f_{W}=J(f,f),
\end{equation*}
where $f=f(r,\varphi,\theta,U,V,W,t)$, and the collision integral
\begin{equation*}
	J(f,f)=\int\limits_{\mathbb{R}^{3}}\int\limits_{S^{2}}{B(g_{s},\theta_{1})(f^{*}f_{1}^{*}-ff_{1})\sin{\theta_{1}}}\,d\theta_{1}\,d\epsilon\,dU_{1}\,dV_{1}\,dW_{1}.
\end{equation*}
Here
\begin{align*}
	&f=f(r,\varphi,\theta,U,V,W,t),~f_{1}=f(r,\varphi,\theta,U_{1},V_{1},W_{1},t),\\
	&f^{*}=f(r,\varphi,\theta,U^{*},V^{*},W^{*},t),~f_{1}^{*}=f(r,\varphi,\theta,U_{1}^{*},V_{1}^{*},W_{1}^{*},t),
\end{align*}
and
\begin{align*}
	&U^{*}=\frac{1}{2}(U+U_{1}+g_{s}n_{1s}),~V^{*}=\frac{1}{2}(V+V_{1}+g_{s}n_{2s}),~W^{*}=\frac{1}{2}(W+W_{1}+g_{s}n_{3s}),\\
	&U_{1}^{*}=\frac{1}{2}(U+U_{1}-g_{s}n_{1s}),~V_{1}^{*}=\frac{1}{2}(V+V_{1}-g_{s}n_{2s}),~W_{1}^{*}=\frac{1}{2}(W+W_{1}-g_{s}n_{3s}),\\
	&\textbf{g}_{s}=((U-U_{1}),(V-V_{1}),(W-W_{1})),~g_{s}=\|\textbf{g}_{s}\|_{2},~\textbf{n}_{s}=(n_{1s},n_{2s},n_{3s}),\\
	&n_{1s}=\cos{(\epsilon-\varphi)}\sin{\theta}\sin{\theta_{1}}+\cos{\theta_{1}}\cos{\theta},~n_{2s}=\cos{(\epsilon-\varphi)}\cos{\theta}\sin{\theta_{1}}-\cos{\theta_{1}}\sin{\theta},\\
	&n_{3s}=\sin{(\epsilon-\varphi)}\sin{\theta_{1}}.
\end{align*}
\textit{Remark:}~$\textbf{n}=(\sin{\theta_{1}}\cos{\epsilon},\sin{\theta_{1}}\sin{\epsilon},\cos{\theta_{1}})$,~$g_{s}=g$, and $\textbf{g}_{s}\cdot\textbf{n}_{s}=\textbf{g}\cdot\textbf{n}$.

\section{Group classification of the Boltzmann equation with a source function}\label{AppendixGroupClassification}
Complete results of the preliminary group classification of the Boltzmann equation with a source function are shown in this Appendix. Numbers in the first column are subalgebra numbers of the form \verb|m.n|, where \verb|n| represents number of a subalgebra from \verb|m|-dimensional subalgebras. The superscripts $^{c}$, and $^{s}$ which are next to this subalgebra number in the first column indicate that the source function $q$ is presented in cylindrical or spherical coordinate systems, respectively. Here $\Psi_{k}$ is an arbitrary function of $k$ independent variables,  and $C$ is constant.

\begin{landscape}
\begin{table}[h]
	\caption{Group classification of Eq. (\ref{eq:fullBoltzannEqWithSources}).}
	\centering
	\begin{tabular}{cll}
		\hline
		Subalgebra no. & Source function $q$ & Subalgebra\\
		\hline
		1.1$^{c}$&$t^{-2}\Psi_{7}(\frac{x}{t}-\frac{\beta}{\alpha}\ln{t},\frac{r}{t},\theta-\frac{1}{\alpha}\ln{t},u-\frac{\beta}{\alpha}\ln{t},V,W,ft)$&$\beta 4+7+\alpha 11$,~$\alpha \neq 0$\\
		1.2$^{c}$&$\Psi_{7}(t,r,\beta\theta-\frac{x}{t},u-\frac{x}{t},V,W,f)$&$\beta 4+7$,~$ \beta \neq 0$\\
		1.3$^{c}$&$\Psi_{7}(t,x,r,u,V,W,f)$&$7$\\
		1.4$^{c}$&$\Psi_{7}(t,r,x-\theta,u,V,W,f)$&$1+7$\\
		1.5$^{c}$&$\Psi_{7}(t^{2}-2x,r,t-\beta\theta,u-t,V,W,f)$&$\beta 4+7+\beta 10$,~$ \beta \neq 0$\\
		1.6$^{c}$&$\Psi_{7}(x,r,t-\theta,u,V,W,f)$&$7+10$\\
		1.7&$t^{-2}\Psi_{7}(\frac{y}{t},\frac{z}{t},\frac{x}{t}-\beta\ln {t},u-\beta\ln {t},v,w,ft)$&$\beta 4+11$,~$\beta \neq 0$\\
		1.8&$t^{-2}\Psi_{7}(\frac{x}{t},\frac{y}{t},\frac{z}{t},u,v,w,ft)$&$11$\\
		1.9&$\Psi_{7}(t^{2}-2x,y,z,u-t,v,w,f)$&$4+10$\\
		1.10&$\Psi_{7}(x,y,z,u,v,w,f)$&$10$\\
		1.11&$\Psi_{7}(t,x-tz,y,u-z,v,w,f)$&$3+4$\\
		1.12&$\Psi_{7}(t,y,z,u-\frac{x}{t},v,w,f)$&$4$\\
		1.13&$\Psi_{7}(t,y,z,u,v,w,f)$&$1$\\
		2.1$^{c}$&$x^{-2}\Psi_{6}(\frac{r}{x},\alpha\theta-\ln{x},u,V,W,fx)$&$10,7+\alpha11$,~$\alpha\neq 0$\\
		2.2$^{c}$&$t^{-2}\Psi_{6}(\frac{r}{t},\frac{x}{t}-\alpha\theta-\beta\ln{t},u-\alpha\theta-\beta\ln{t},V,W,ft)$&$\alpha4+7,\beta4+11$\\
		2.3$^{c}$&$t^{-2}\Psi_{6}(\frac{r}{t},\alpha\theta-\ln{t},u-\frac{x}{t},V,W,ft)$&$4,7+\alpha11$,~$\alpha\neq 0$\\
		2.4$^{c}$&$t^{-2}\Psi_{6}(\frac{r}{t},\alpha\theta-\ln{t},u-\frac{\beta}{\alpha}\ln{t},V,W,ft)$&$1,\beta4+7+\alpha11$,~$\alpha\neq 0$\\
		2.5$^{c}$&$\Psi_{6}(r,x,u,V,W,f)$&$7,10$\\
		2.6$^{c}$&$\Psi_{6}(r,x-\theta,u,V,W,f)$&$1+7,10$\\
		2.7$^{c}$&$\Psi_{6}(r,2(x-\alpha\theta)-t^{2},u-t,V,W,f)$&$\alpha1+7,4+10$\\
		2.8$^{c}$&$\Psi_{6}(t,r,u-\frac{x}{t},V,W,f)$&$4,7$\\
		2.9$^{c}$&$\Psi_{6}(t,r,u-\beta\theta,V,W,f)$&$1,\beta4+7$\\
		2.10$^{c}$&$\Psi_{6}(t,r,u-\frac{x}{t}+\frac{\theta}{t},V,W,f)$&$4,1+7$\\
		2.11$^{c}$&$\Psi_{6}(r,\theta-t,u-\beta t,V,W,f)$&$1,\beta4+7+10$\\
		2.12&$x^{-2}\Psi_{6}(\frac{y}{x},\frac{z}{x},u,v,w,fx)$&$10,11$\\
		2.13&$t^{-2}\Psi_{6}(\frac{y}{t},\frac{z}{t},u-\frac{x}{t},v,w,ft)$&$4,11$\\
		2.14&$t^{-2}\Psi_{6}(\frac{y}{t}-\alpha\ln{t},\frac{z}{t},u-\frac{x}{t},v-\alpha\ln{t},w,ft)$&$4,\alpha5+11$,~$\alpha\neq 0$\\
		2.15&$t^{-2}\Psi_{6}(\frac{z}{t},\frac{y}{t}-\alpha\ln{t},u-\beta\ln{t},v-\alpha\ln{t},w,ft)$&$1,\beta4+\alpha5+11$,~$\alpha\neq 0$\\
		2.16&$t^{-2}\Psi_{6}(\frac{y}{t},\frac{z}{t},u-\beta\ln{t},v,w,ft)$&$1,\beta4+11$\\
		2.17&$\Psi_{6}(y,z,u,v,w,f)$&$1,10$\\
		2.18&$\Psi_{6}(2x-t^{2},y,u-t,v,w-\alpha t,f)$&$3,4+\alpha6+10$,~$\alpha\neq 0$\\
		2.19&$\Psi_{6}(y,z,u-t,v,w,f)$&$1,4+10$\\
		\hline
	\end{tabular}
\end{table}
\end{landscape}

\newpage
\begin{landscape}
	\begin{table}[h]
		\addtocounter{table}{-1}
		\caption{Group classification of Eq. (\ref{eq:fullBoltzannEqWithSources}) (Continued).}
		\centering
		\begin{tabular}{cll}
			\hline
			Subalgebra no. & Source function $q$ & Subalgebra\\
			\hline

			2.20&$\Psi_{6}(t,\alpha(ty-\tau z)+\beta(tz-\sigma y)+x(\sigma\tau-t^{2}),u,v+\frac{\beta y-\tau x}{\alpha\tau-\beta t},w+\frac{tx-\alpha y}{\alpha\tau-\beta t},f)$&$\alpha1+\sigma3+5,\beta1+\tau2+6$,\\
			& &$\alpha ^2+\beta ^2 +(\sigma+\tau)^2=1$\\
			2.21&$\Psi_{6}(t,x,u,v-\frac{z+ty}{t^2+1},w+\frac{y-tz}{t^2+1},f)$&$3+5,2-6$\\
			2.22&$\Psi_{6}(t,x,u,v-\frac{y}{t},w-\frac{z}{t},f)$&$5,6$\\
			2.23&$\Psi_{6}(t,x-\alpha y-tz,u-z,v,w,f)$&$\alpha1+2,3+4$\\
			2.24&$\Psi_{6}(t,z,u+\frac{\alpha y-x}{t},v,w,f)$&$\alpha1+2,4$\\
			2.25&$\Psi_{6}(t,y,u-z,v,w,f)$&$1,3+4$\\
			2.26&$\Psi_{6}(t,y,z,v,w,f)$&$1,4$\\
			2.27&$\Psi_{6}(t,x,u,v,w,f)$&$2,3$\\
			3.1$^{c}$&$x^{-2}\Psi_{5}(\frac{r}{x},u,V,W,fx)$&$7,10,11$\\
			3.2$^{c}$&$e^{-2\alpha\theta}\Psi_{5}(re^{-\alpha\theta},u-\beta\theta,V,W,fe^{\alpha\theta})$&$1,10,\beta4+7+\alpha11$\\
			3.3$^{c}$&$t^{-2}\Psi_{5}(\frac{r}{t},u-\frac{x}{t},V,W,ft)$&$4,7,11$\\
			3.4$^{c}$&$t^{-2}\Psi_{5}(\frac{r}{t},u-\alpha\theta-\beta\ln{t},V,W,ft)$&$1,\alpha4+7,\beta4+11$\\
			3.5&$t^{-2}\Psi_{5}(\frac{x}{t}-\frac{\beta}{\alpha}\ln{t},u-\frac{\beta}{\alpha}\ln{t},\arctan(\frac{w-\frac{z}{t}}{v-\frac{y}{t}})-\frac{1}{\alpha}\ln{t},\sqrt{(v-\frac{y}{t})^{2}+(w-\frac{z}{t})^{2}},ft)$&$5,6,\beta4+7+\alpha11$,~$\alpha\neq0$\\
			3.6$^{c}$&$t^{-2}\Psi_{5}(\frac{r}{t},\alpha\theta-\ln{t},V,W,ft)$&$1,4,7+\alpha11$,~$\alpha\neq 0$\\
			3.7&$t^{-2}\Psi_{5}(\frac{x}{t}-\frac{\beta}{\alpha}\ln{t},u-\frac{\beta}{\alpha}\ln{t},\arctan(\frac{w}{v})-\frac{1}{\alpha}\ln{t},\sqrt{v^{2}+w^{2}},ft)$&$2,3,\beta4+7+\alpha11$,~$\alpha\neq 0$\\
			3.8$^{s}$&$\Psi_{5}(t,r,U,V^{2}+W^{2},f)$&$7,8,9$\\
			3.9$^{c}$&$\Psi_{5}(r,u-t-\alpha\theta,V,W,f)$&$1,\alpha4+7,4+10$\\
			3.10&$\Psi_{5}(t,\frac{x}{t}-\beta\arctan(\frac{w-\frac{z}{t}}{v-\frac{y}{t}}),u-\beta\arctan(\frac{w-\frac{z}{t}}{v-\frac{y}{t}}),\sqrt{(v-\frac{y}{t})^{2}+(w-\frac{z}{t})^{2}},f)$&$5,6,\beta4+7$\\
			3.11$^{c}$&$\Psi_{5}(t,r,V,W,f)$&$1,4,7$\\
			3.12&$\Psi_{5}(t,u-\frac{x}{t},\arctan(\frac{w}{v})-\frac{x}{\beta t},\sqrt{v^{2}+w^{2}},f)$&$2,3,\beta4+7$,~$\beta\neq 0$\\
			3.13&$\Psi_{5}(t,x,u,\sqrt{v^{2}+w^{2}},f)$&$2,3,7$\\
			3.14&$\Psi_{5}(t,u-\frac{\alpha x}{1+\alpha t},\arctan(\frac{w-\frac{z}{t}}{v-\frac{y}{t}})-\frac{x}{1+\alpha t},\sqrt{(v-\frac{y}{t})^{2}+(w-\frac{z}{t})^{2}},f)$&$5,6,1+\alpha4+7$\\
			3.15&$\Psi_{5}(t,u+\beta\arctan(\frac{y-tv+w}{z-tw-v}),x+(\alpha+\beta t)\arctan(\frac{y-tv+w}{z-tw-v}),$&$3+5,2-6,\alpha1+\beta4+7$\\
			&$\sqrt{(y-tv+w)^{2}+(z-tw-v)^{2}},f)$&\\
			3.16&$\Psi_{5}(t,u,\arctan(\frac{w}{v})-x,\sqrt{v^{2}+w^{2}},f)$&$2,3,1+7$\\
			3.17$^{c}$&$\Psi_{5}(t-\theta,r,V,W,f)$&$1,4,7+10$\\
			3.18&$\Psi_{5}(2x-t^{2},u-t,\arctan(\frac{w}{v})-\frac{t}{\beta},\sqrt{v^{2}+w^{2}},f)$&$2,3,\beta4+7+\beta10$,~$\beta\neq 0$\\
			\hline
		\end{tabular}
	\end{table}
\end{landscape}

\newpage
\begin{landscape}
	\begin{table}[h]
		\addtocounter{table}{-1}
		\caption{Group classification of Eq. (\ref{eq:fullBoltzannEqWithSources}) (Continued).}
		\centering
		\begin{tabular}{cll}
			\hline
			Subalgebra no. & Source function $q$ & Subalgebra\\
			\hline

			3.19&$\Psi_{5}(x,u,\arctan(\frac{w}{v})-t,\sqrt{v^{2}+w^{2}},f)$&$2,3,7+10$\\
			3.20&$y^{-2}\Psi_{5}(\frac{y}{z},u-\beta\ln{y},v,w,fy)$&$1,10,\beta4+11$\\
			3.21&$t^{-2}\Psi_{5}(\frac{x}{t}-\beta\ln{t},u-\beta\ln{t},v-\frac{y}{t},w-\frac{z}{t},ft)$&$5,6,\beta4+11$\\
			3.22&$t^{-2}\Psi_{5}(\frac{y}{t}-\sigma\ln{t},u-\frac{\alpha z}{t}-\beta\ln{t},v-\sigma\ln{t},w-\frac{z}{t},ft)$&$1,\alpha4+6,\beta4+\sigma5+11$\\
			3.23&$t^{-2}\Psi_{5}(\frac{y}{t},\frac{z}{t}-\sigma\ln{t},v,w-\sigma\ln{t},ft)$&$1,4,\sigma6+11$,~$\sigma\neq 0$\\
			3.24&$t^{-2}\Psi_{5}(\frac{y}{t},\frac{z}{t},v,w,ft)$&$1,4,11$\\
			3.25&$t^{-2}\Psi_{5}(\frac{x}{t}-\beta\ln{t},u-\beta\ln{t},v-\sigma\ln{t},w,ft)$&$2,3,\beta4+\sigma5+11$,~$\sigma\neq 0$\\
			3.26&$t^{-2}\Psi_{5}(\frac{x}{t}-\beta\ln{t},u-\beta\ln{t},v,w,ft)$,&$2,3,\beta 4+11$\\
			3.27&$\Psi_{5}(t^{2}-2x+2\alpha w,y-\beta w,u-t,v,f)$&$3,\alpha1+\beta2+6,4+10$\\
			3.28&$\Psi_{5}(z,u-y,v,w,f)$&$1,2+4,10$\\
			3.29&$\Psi_{5}(y,z,v,w,f)$&$1,4,10$\\
			3.30&$\Psi_{5}(t^{2}-2x,u-t,v,w-\sigma t,f)$&$2,3,4+\sigma6+10$,~$\sigma\neq 0$\\
			3.31&$\Psi_{5}(t^{2}-2x,u-t,v,w,f)$&$2,3,4+10$\\
			3.32&$\Psi_{5}(x,u,v,w-t,f)$&$2,3,6+10$\\
			3.33&$\Psi_{5}(x,u,v,w,f)$&$2,3,10$\\
			3.34&$\Psi_{5}(t,x-tu-\delta v+\beta w,y-tv+\delta u-\sigma v-\alpha w,$&$-\delta2+\beta3+4,\delta1+\sigma2-\alpha3+5,-\beta1+\alpha2+\tau3+6$,\\
			&$z-tw-\beta u+\alpha v-\tau w,f)$&$\alpha^2+\beta^2+\delta^2+(\sigma+\tau)^2=1$\\
			3.35&$\Psi_{5}(t,x-tu,tw-z+v,y-tz+(t^{2}+1)w,f)$&$4,3+5,2-6$\\
			3.36&$\Psi_{5}(t,u(t+1)-x,y-tv,z-tw,f)$&$1+4,5,6$\\
			3.37&$\Psi_{5}(t,u-\frac{x}{t},v-\frac{y}{t},w-\frac{z}{t},f)$&$4,5,6$\\
			3.38&$\Psi_{5}(t,\tau w+tv-y,w(\sigma-\alpha t)+\beta v-x+\alpha z,u,f)$&$\alpha1+3,\beta1+5,\sigma1+\tau2+6$,~$\beta^2+\sigma^2+\tau^2=1$\\
			3.39&$\Psi_{5}(t,y-tv,x-\alpha (z-tw),u,f)$&$\alpha1+3,5,6$\\
			3.40&$\Psi_{5}(t,w(t^{2}-\tau)+y-tz,\tau w-y+tv,u,f)$&$1,3+5,\tau2+6$,~$\tau\neq -1$\\
			3.41&$\Psi_{5}(t,w(t^{2}+1)+y-tz,w+y-tv,u,f)$&$1,3+5,2-6$\\
			3.42&$\Psi_{5}(t,u,v-\frac{y}{t},w-\frac{z}{t},f)$&$1,5,6$\\
			3.43&$\Psi_{5}(t,tu-x+\beta z,v,w,f)$&$\beta1+3,2,4$\\
			3.44&$\Psi_{5}(t,u-\frac{x}{t},v,w,f)$&$2,3,4$\\
			3.45&$\Psi_{5}(t,u-z,v,w,f)$&$1,2,3+4$\\
			3.46&$\Psi_{5}(t,z,v,w,f)$&$1,2,4$\\
			3.47&$\Psi_{5}(t,u,v,w,f)$&$1,2,3$\\
			\hline
		\end{tabular}
	\end{table}
\end{landscape}

\newpage
\begin{landscape}
	\begin{table}[h]
		\addtocounter{table}{-1}
		\caption{Group classification of Eq. (\ref{eq:fullBoltzannEqWithSources}) (Continued).}
		\centering
		\begin{tabular}{cll}
			\hline
			Subalgebra no. & Source function $q$  & Subalgebra\\
			\hline
			
			4.1$^{s}$&$r^{-2}\Psi_{4}(\frac{t}{r},U,V^{2}+W^{2},fr)$&$7,8,9,11$\\
			4.2$^{c}$&$r^{-2}\Psi_{4}(u-\alpha\theta,V,W,fr)$&$1,\alpha4+7,10,11$\\
			4.3&$e^{2\alpha\arctan(\frac{v}{w})}\Psi_{4}(xe^{\alpha\arctan(\frac{v}{w})},\sqrt{v^{2}+w^{2}},u,fe^{-\alpha\arctan(\frac{v}{w})})$&$2,3,10,7+\alpha11$\\
			4.4$^{c}$&$r^{-2}\Psi_{4}(\alpha\theta-\ln{r},V,W,fr)$&$1,4,10,7+\alpha11$,~$\alpha\neq 0$\\
			4.5&$t^{-2}\Psi_{4}(\frac{\sqrt{(y-tv)^{2}+(z-tw)^{2}}}{t},t^{-\beta}e^{\frac{x-\alpha t \arctan(\frac{z-tw}{y-tv})}{t}},t^{-\beta}e^{u-\alpha\arctan(\frac{z-tw}{y-tv})},ft)$&$5,6,\alpha4+7,\beta4+11$\\
			4.6$^{c}$&$r^{-2}\Psi_{4}(\frac{t}{r},V,W,fr)$&$1,4,7,11$\\
			4.7&$t^{-2}\Psi_{4}(t^{-\beta}e^{\frac{x+\alpha t\arctan(\frac{v}{w})}{t}},t^{-\beta}e^{u+\alpha\arctan(\frac{v}{w})},\sqrt{v^{2}+w^{2}},ft)$&$2,3,\alpha4+7,\beta4+11$\\
			4.8&$t^{-2}\Psi_{4}(\frac{x-tu}{t},\frac{\sqrt{(y-tv)^{2}+(z-tw)^{2}}}{t},te^{-\alpha\arctan(\frac{z-tw}{y-tv})},ft)$&$4,5,6,7+\alpha11$,~$\alpha\neq 0$\\
			4.9&$e^{-2\alpha\arctan(\frac{z-tw}{y-tv})}\Psi_{4}(\sqrt{(y-tv)^{2}+(z-tw)^{2}}e^{-\alpha\arctan(\frac{z-tw}{y-tv})},te^{-\alpha\arctan(\frac{z-tw}{y-tv})},$&$1,5,6,\beta4+7+\alpha11$\\
			&$u-\beta\arctan(\frac{z-tw}{y-tv}),fe^{\alpha\arctan(\frac{z-tw}{y-tv})})$&\\
			4.10&$e^{2\alpha\arctan(\frac{v}{w})}\Psi_{4}((x-tu)e^{\alpha\arctan(\frac{v}{w})},te^{\alpha\arctan(\frac{v}{w})},\sqrt{v^{2}+w^{2}},fe^{-\alpha\arctan(\frac{v}{w})})$&$2,3,4,7+\alpha11$\\
			4.11&$t^{-2}\Psi_{4}(te^{\alpha\arctan(\frac{v}{w})},t^{-\beta}e^{\alpha u},\sqrt{v^{2}+w^{2}},ft)$&$1,2,3,\beta4+7+\alpha11$,~$\alpha\neq 0$\\
			4.12&$\Psi_{4}(u+\beta\arctan(\frac{v}{w}),\sqrt{v^{2}+w^{2}},t,f)$&$1,2,3,\beta4+7$\\
			4.13$^{s}$&$\Psi_{4}(r,U,V^{2}+W^{2},f)$&$7,8,9,10$\\
			4.14&$\Psi_{4}(v^{2}+w^{2},x,u,f)$&$2,3,7,10$\\
			4.15&$\Psi_{4}(x+\arctan(\frac{v}{w}),\sqrt{v^{2}+w^{2}},u,f)$&$2,3,1+7,10$\\
			4.16&$\Psi_{4}(2\alpha\arctan(\frac{v}{w})+2x-t^{2},\sqrt{v^{2}+w^{2}},t-u,f)$&$2,3,\alpha1+7,4+10$\\
			4.17&$\Psi_{4}(x-tu,(y-tv)^{2}+(z-tw)^{2},t,f)$&$4,5,6,7$\\
			4.18&$\Psi_{4}(x-tu-\arctan(\frac{z-tw}{y-tv}),\sqrt{(y-tv)^{2}+(z-tw)^{2}},t,f)$&$4,5,6,1+7$\\
			4.19&$\Psi_{4}(x-tu-\alpha\arctan(\frac{z-v-tw}{w+y-tv}),\sqrt{(w+y-tv)^{2}+(z-v-tw)^{2}},t,f)$&$4,3+5,2-6,\alpha1+7$\\
			4.20&$\Psi_{4}(u-\alpha\arctan(\frac{z-v-tw}{w+y-tv}),\sqrt{(w+y-tv)^{2}+(z-v-tw)^{2}},t,f)$&$1,3+5,2-6,\alpha4+7$\\
			4.21&$\Psi_{4}(x-tu+\arctan(\frac{v}{w}),\sqrt{v^{2}+w^{2}},t,f)$&$2,3,4,1+7$\\
			4.22&$\Psi_{4}(u-\beta t,t+\arctan(\frac{v}{w}),\sqrt{v^{2}+w^{2}},f)$&$1,2,3,\beta4+7+10$\\
			4.23&$y^{-2}\Psi_{4}(\frac{z}{y},v,w,fy)$&$1,4,10,11$\\
			4.24&$x^{-2}\Psi_{4}(u,v,w-\alpha\ln{x},fx)$&$2,3,10,\alpha6+11$,~$\alpha\neq 0$\\
			4.25&$x^{-2}\Psi_{4}(u,v,w,fx)$&$2,3,10,11$\\
			4.26&$t^{-2}\Psi_{4}(u-\frac{x}{t},v-\frac{y}{t},w-\frac{z}{t},ft)$&$4,5,6,11$\\
			4.27&$t^{-2}\Psi_{4}(\frac{y-tv}{t},\frac{z-tw}{t},t^{-\beta}e^{u-\alpha v},ft)$&$1,\alpha4+5,6,\beta4+11$,~$\alpha\neq 0$\\
			4.28&$t^{-2}\Psi_{4}(u-\beta\ln{t},v-\frac{y}{t},w-\frac{z}{t},ft)$&$1,5,6,\beta4+11$\\
			
			\hline
		\end{tabular}
	\end{table}	
\end{landscape}

\newpage
\begin{landscape}
	\begin{table}[h!]
		\addtocounter{table}{-1}
		\caption{Group classification of Eq. (\ref{eq:fullBoltzannEqWithSources}) (Continued).}
		\centering
		\begin{tabular}{cll}
			\hline
			Subalgebra no. & Source function $q$ & Subalgebra\\
			\hline
			4.29&$t^{-2}\Psi_{4}(\frac{y}{t}-\alpha\ln{t},v-\alpha\ln{t},w-\frac{z}{t},ft)$&$1,4,6,\alpha5+11$\\
			4.30&$t^{-2}\Psi_{4}(t^{-\beta}e^{\frac{x-\alpha tw}{t}},t^{-\beta}e^{u-\alpha w},t^{-\sigma}e^{v},ft)$&$2,3,\alpha4+6,\beta4+\sigma5+11$\\
			4.31&$t^{-2}\Psi_{4}(u-\frac{x}{t},v-\alpha\ln{t},w-\beta\ln{t},ft)$&$2,3,4,\alpha5+\beta6+11$,~$\alpha^2+\beta^2\neq0$\\
			4.32&$t^{-2}\Psi_{4}(u-\frac{x}{t},v,w,ft)$&$2,3,4,11$\\
			4.33&$t^{-2}\Psi_{4}(u-\beta\ln{t},v,w,ft)$&$1,2,3,\beta4+11$,~$\beta\neq 0$\\
			4.34&$t^{-2}\Psi_{4}(u,v,w,ft)$&$1,2,3,11$\\
			4.35&$\Psi_{4}(t^{2}-2x+2\alpha v,u-t,w-\beta t,f)$&$2,3,\alpha1+5,4+\beta6+10$\\
			4.36&$\Psi_{4}(x-\alpha v,u,w-t,f)$&$2,3,\alpha1+5, 6+10$\\
			4.37&$\Psi_{4}(u,v-x,w,f)$&$2,3,1+5,10$\\
			4.38&$\Psi_{4}(x,u,w,f)$&$2,3,5,10$\\
			4.39&$\Psi_{4}(u-t,v,w,f)$&$1,2,3,4+10$\\
			4.40&$\Psi_{4}(u,v,w,f)$&$1,2,3,10$\\
			4.41&$\Psi_{4}(y-\sigma u-tv-\beta w,z-\tau u-\alpha v-tw,t,f)$&$1,\sigma2+\tau3+4,\alpha3+5,\beta2+6$,\\
			& &$\sigma^2+\tau^2+(\alpha+\beta)^2=1$\\
			4.42&$\Psi_{4}(y-tv+w,y-tz+w(t^{2}+1),t,f)$&$1,4,3+5,2-6$\\
			4.43&$\Psi_{4}(v-\frac{y}{t},w-\frac{z}{t},t,f)$&$1,4,5,6$\\
			4.44&$\Psi_{4}(v-x+\alpha (z-tw),u,t,f)$&$2,\alpha1+3,1+5,6$,~$\alpha\neq 0$\\
			4.45&$\Psi_{4}(u,v-x,t,f)$&$2,3,1+5,6$\\
			4.46&$\Psi_{4}(t,u,w+\frac{\beta x-z}{t},f)$&$1+\beta3,2,5,6$\\
			4.47&$\Psi_{4}(x,u,t,f)$&$2,3,5,6$\\
			4.48&$\Psi_{4}(v-z+tw,u,t,f)$&$1,2,3+5,6$\\
			4.49&$\Psi_{4}(u,w-\frac{z}{t},t,f)$&$1,2,5,6$\\
			4.50&$\Psi_{4}(v,w,t,f)$&$1,2,3,4$\\
			5.1$^{s}$&$r^{-2}\Psi_{3}(U,V^{2}+W^{2},fr)$&$7,8,9,10,11$\\
			5.2$^{c}$&$r^{-2}\Psi_{3}(V,W,fr)$&$1,4,7,10,11$\\
			5.3&$x^{-2}\Psi_{3}(v^{2}+w^{2},u,fx)$&$2,3,7,10,11$\\
			5.4&$e^{2\alpha\arctan(\frac{v}{w})}\Psi_{3}(u+\beta\arctan(\frac{v}{w}),\sqrt{v^{2}+w^{2}},fe^{-\alpha\arctan(\frac{v}{w})})$&$1,2,3,10,\beta4+7+\alpha11$\\
			5.5&$t^{-2}\Psi_{3}(\frac{x-tu}{t},\frac{(y-tv)^{2}+(z-tw)^{2}}{t^{2}},ft)$&$4,5,6,7,11$\\
			5.6&$t^{-2}\Psi_{3}(\frac{x-tu}{t},v^{2}+w^{2},ft)$&$2,3,4,7,11$\\
			5.7&$t^{-2}\Psi_{3}(t^{\beta}e^{\alpha\arctan(\frac{z-tw}{y-tv})-u},\sqrt{(\frac{y-tv}{t})^{2}+(\frac{z-tw}{t})^{2}},ft)$&$1,5,6,\alpha4+7,\beta4+11$\\
			5.8&$t^{-2}\Psi_{3}(t^{-\beta}e^{u+\alpha\arctan(\frac{v}{w})},\sqrt{v^{2}+w^{2}},ft)$&$1,2,3,\alpha4+7,\beta4+11$\\
			
			\hline
		\end{tabular}
	\end{table}	
\end{landscape}

\newpage
\begin{landscape}
	\begin{table}[h]
		\addtocounter{table}{-1}
		\caption{Group classification of Eq. (\ref{eq:fullBoltzannEqWithSources}) (Continued).}
		\centering
		\begin{tabular}{cll}
			\hline
			Subalgebra no. & Source function $q$ & Subalgebra\\
			\hline
			5.9&$e^{-2\alpha\arctan(\frac{z-tw}{y-tv})}\Psi_{3}(\sqrt{(y-tv)^{2}+(z-tw)^{2}}e^{-\alpha\arctan(\frac{z-tw}{y-tv})},te^{-\alpha\arctan(\frac{z-tw}{y-tv})},$&$1,4,5,6,7+\alpha11$\\
			&$fe^{\alpha\arctan(\frac{z-tw}{y-tv})})$&\\
			5.10&$t^{-2}\Psi_{3}(\frac{x}{t}-\frac{\beta}{\alpha}\ln{t},u-\frac{\beta}{\alpha}\ln{t},ft)$&$2,3,5,6,\beta4+7+\alpha11$,~$\alpha\neq 0$\\
			5.11&$t^{-2}\Psi_{3}(te^{\alpha\arctan(\frac{v}{w})},\sqrt{v^{2}+w^{2}},ft)$&$1,2,3,4,7+\alpha11$,~$\alpha\neq 0$\\
			5.12&$\Psi_{3}(t-u-\alpha\arctan(\frac{v}{w}),\sqrt{v^{2}+w^{2}},f)$&$1,2,3,\alpha4+7,4+10$\\
			5.13&$\Psi_{3}(u-\frac{x}{t},t,f)$&$2,3,5,6,\beta4+7$,~$\beta\neq 0$\\
			5.14&$(4.47)$&$2,3,5,6,7$\\
			5.15&$\Psi_{3}(v^{2}+w^{2},t,f)$&$1,2,3,4,7$\\
			5.16&$\Psi_{3}((w+y-tv)^{2}+(v-z+tw)^{2},t,f)$&$1,4,3+5,2-6,7$\\
			5.17&$\Psi_{3}(u,t,f)$&$2,3,5,6,1+7$\\
			5.18&$\Psi_{3}(t^{2}-2x,u-t,f)$&$2,3,5,6,\beta4+7+\beta10$,~$\beta\neq 0$\\
			5.19&$\Psi_{3}(x,u,f)$&$2,3,5,6,7+10$\\
			5.20&$\Psi_{3}(t+\arctan(\frac{v}{w}),\sqrt{v^{2}+w^{2}},f)$&$1,2,3,4,7+10$\\
			5.21&$x^{-2}\Psi_{3}(u,w-\beta\ln{x},fx)$&$2,3,5,10,\beta6+11$\\
			5.22&$e^{-2\beta u}\Psi_{3}(v,w,fe^{\beta u})$&$1,2,3,10,4+\beta11$\\
			5.23&$f^{2}\Psi_{3}(u,v,w)$&$1,2,3,10,11$\\
			5.24&$t^{-2}\Psi_{3}(v-\frac{y}{t},w-\frac{z}{t},ft)$&$1,4,5,6,11$\\
			5.25&$t^{-2}\Psi_{3}(u-\frac{x}{t},v-\frac{x}{\alpha t}+\frac{\beta}{\alpha}\ln{t},ft)$&$2,3,\alpha4+5,6,\beta4+11$,~$\alpha\neq 0$\\
			5.26&$t^{-2}\Psi_{3}(\frac{x}{t}-\beta\ln{t},u-\beta\ln{t},ft)$&$2,3,5,6,\beta4+11$\\
			5.27&$t^{-2}\Psi_{3}(u-\frac{x}{t},v-\beta\ln{t},ft)$&$2,3,4,6,\beta5+11$\\
			5.28&$t^{-2}\Psi_{3}(u-\beta\ln{t},v,ft)$&$1,2,3,6,\beta4+11$,~$\beta\neq 0$\\
			5.29&$t^{-2}\Psi_{3}(v,w,ft)$&$1,2,3,4,11$\\
			5.30&$\Psi_{3}(2(x-\alpha v)-t^{2},u-t,f)$&$2,3,\alpha1+5,6,4+10$,~$\alpha\neq 0$\\
			5.31&$\Psi_{3}(t^{2}-2x,u-t,f)$&$2,3,5,6,4+10$\\
			5.32&$\Psi_{3}(u,v-x,f)$&$2,3,1+5,6,10$\\
			5.33&$\Psi_{3}(x,u,f)$&$2,3,5,6,10$\\
			5.34&$\Psi_{3}(u-t,v,f)$&$1,2,3,6,4+10$\\
			5.35&$\Psi_{3}(u-\frac{x}{t},t,f)$&$2,3,4,5,6$\\
			5.36&$\Psi_{3}(t,tu-x+w,f)$&$2,3,4,5,1+6$\\
			5.37&$\Psi_{3}(u,t,f)$&$1,2,3,5,6$\\
			6.1&$f^{2}\Psi_{2}(v^{2}+w^{2},u)$&$1,2,3,7,10,11$\\
			6.2&$x^{-2}\Psi_{2}(u,fx)$&$2,3,5,6,10,7+\alpha11$,~$\alpha\neq 0$\\
			
			\hline
		\end{tabular}
	\end{table}
\end{landscape}

\newpage
\begin{landscape}
	\begin{table}[h]
		\addtocounter{table}{-1}
		\caption{Group classification of Eq. (\ref{eq:fullBoltzannEqWithSources}) (Continued).}
		\centering
		\begin{tabular}{cll}
			\hline
			Subalgebra no. & Source function $q$ & Subalgebra\\
			\hline
			
			6.3&$e^{2\alpha\arctan(\frac{v}{w})}\Psi_{2}(\sqrt{v^{2}+w^{2}},fe^{-\alpha\arctan(\frac{v}{w})})$&$1,2,3,4,10,7+\alpha11$\\
			6.4&$t^{-2}\Psi_{2}((v-\frac{y}{t})^{2}+(w-\frac{z}{t})^{2},ft)$&$1,4,5,6,7,11$\\
			6.5&$t^{-2}\Psi_{2}(v^{2}+w^{2},ft)$&$1,2,3,4,7,11$\\
			6.6&$t^{-2}\Psi_{2}(u-\frac{x}{t},ft),~\alpha\neq0$&$2,3,5,6,\alpha4+7,\beta4+11$\\
			&$(5.26),~\alpha=0$&\\
			6.7&$t^{-2}\Psi_{2}(\frac{x-tu}{t},ft)$&$2,3,4,5,6,7+\alpha11$,~$\alpha\neq 0$\\
			6.8&$e^{-\frac{2\alpha u}{\beta}}\Psi_{2}(te^{-\frac{\alpha u}{\beta}},fe^{\frac{\alpha u}{\beta}}), ~\beta\neq0$&$1,2,3,5,6,\beta4+7+\alpha11$\\
			&$t^{-2}\Psi_{2}(u,ft),~\beta=0,\alpha\neq0$&\\
			&$(5.37),~\beta=0,\alpha=0$&\\
			6.9&$\Psi_{3}((u-\frac{x}{t})^{2}+(v-\frac{y}{t})^{2}+(w-\frac{z}{t})^{2},t,f)$&$4,5,6,7,8,9$\\
			6.10&$\Psi_{3}(\sqrt{u^{2}+v^{2}+w^{2}},t,f)$&$1,2,3,7,8,9$\\
			6.11&$\Psi_{2}(u,f)$&$2,3,5,6,1+7,10$\\
			6.12&$\Psi_{2}(u-t,f),~\alpha\neq0$&$2,3,5,6,\alpha1+7,4+10$\\
			&$(5.31),~\alpha=0$&\\
			6.13&$(5.33)$&$2,3,5,6,7,10$\\
			6.14&$\Psi_{2}(t,f)$&$2,3,4,5,6,1+7$\\
			6.15&$(5.35)$&$2,3,4,5,6,7$\\
			6.16&$\Psi_{2}(u,f)$&$1,2,3,5,6,7+10$\\
			6.17&$x^{-2}\Psi_{2}(u,fx)$&$2,3,5,6,10,11$\\
			6.18&$e^{-\frac{2w}{\alpha}}\Psi_{2}(v,fe^{\frac{w}{\alpha}})$&$1,2,3,4,10,\alpha6+11$,~$\alpha\neq 0$\\
			6.19&$f^{2}\Psi_{2}(v,w)$&$1,2,3,4,10,11$\\
			6.20&$t^{-2}\Psi_{2}(u-\alpha\ln{t},ft)$&$1,2,3,5,6,\alpha4+11$\\
			6.21&$t^{-2}\Psi_{2}(\frac{x-tu}{t},ft)$&$2,3,4,5,6,11$\\
			6.22&$\Psi_{2}(u,f)$&$1,2,3,5,6,10$\\
			6.23&$\Psi_{2}(u-t,f)$&$1,2,3,5,6,4+10$\\
			6.24&$\Psi_{2}(t,f)$&$1,2,3,4,5,6$\\
			6.25&$\Psi_{2}(t,f),~\beta\neq 0$&$1,2,3,5,6,\beta4+7$\\
			&$(5.37),~\beta=0$&\\
			7.1&$t^{-2}\Psi_{2}(u^{2}+v^{2}+w^{2},ft)$&$1,2,3,7,8,9,11$\\
			7.2&$t^{-2}\Psi_{2}((u-\frac{x}{t})^{2}+(v-\frac{y}{t})^{2}+(w-\frac{z}{t})^{2},ft)$&$4,5,6,7,8,9,11$\\
			
			\hline
		\end{tabular}
	\end{table}
\end{landscape}

\newpage
\begin{landscape}
	\begin{table}[h]
		\addtocounter{table}{-1}
		\caption{Group classification of Eq. (\ref{eq:fullBoltzannEqWithSources}) (Continued).}
		\centering
		\begin{tabular}{cll}
			\hline
			Subalgebra no. & Source function $q$ & Subalgebra\\
			\hline
			7.3&$(6.17)$&$2,3,5,6,7,10,11$\\
			7.4&$f^{2}\Psi_{1}(v^{2}+w^{2})$&$1,2,3,4,7,10,11$\\
			7.5&$e^{-\frac{2\alpha u}{\beta}}\Psi_{1}(fe^{\frac{\alpha u}{\beta}}), ~\beta\neq0$&$1,2,3,5,6,10,\beta4+7+\alpha11$\\
			&$f^{2}\Psi_{1}(u),~\beta=0,\alpha\neq0$&\\
			&$(6.22),~\beta=0,\alpha=0$&\\
			7.6&$(6.21)$&$2,3,4,5,6,7,11$\\
			7.7&$t^{-2}\Psi_{1}(ft),~\alpha\neq0$&$1,2,3,5,6,\alpha4+7,\beta4+11$\\
			&$(6.20),~\alpha=0$&\\
			7.8&$t^{-2}\Psi_{1}(ft)$&$1,2,3,4,5,6,7+\alpha11$,~$\alpha\neq 0$\\
			7.9&$\Psi_{2}(u^{2}+v^{2}+w^{2},f)$&$1,2,3,7,8,9,10$\\
			7.10&$\Psi_{1}(f),~\alpha\neq0$&$1,2,3,5,6,\alpha4+7,4+10$\\
			&$(6.23),~\alpha=0$&\\
			7.11&$\Psi_{1}(f)$&$1,2,3,4,5,6,7+10$\\
			7.12&$f^{2}\Psi_{1}(f^{\alpha}e^{u})$&$1,2,3,5,6,10,\alpha4+11$\\
			7.13&$t^{-2}\Psi_{1}(ft)$&$1,2,3,4,5,6,11$\\
			7.14&$\Psi_{1}(f)$&$1,2,3,4,5,6,10$\\
			8.1&$f^{2}\Psi_{1}(u^{2}+v^{2}+w^{2})$&$1,2,3,7,8,9,10,11$\\
			8.2&$Cf^{2},~\alpha\neq0$&$1,2,3,5,6,\alpha4+7,10,\beta4+11$\\
			&$(7.12),~\alpha=0$&\\
			8.3&$Cf^{2},~\alpha\neq0$&$1,2,3,4,5,6,7+\alpha11,10$\\
			&$(7.14),~\alpha=0$&\\
			8.4&$(7.13)$&$1,2,3,4,5,6,7,11$\\
			8.5&$Cf^{2}$&$1,2,3,4,5,6,10,11$\\
			9.1&$(8.5)$&$1,2,3,4,5,6,7,10,11$\\
			9.2&$(6.24)$&$1,2,3,4,5,6,7,8,9$\\
			10.1&$(7.13)$&$1,2,3,4,5,6,7,8,9,11$\\
			10.2&$(7.14)$&$1,2,3,4,5,6,7,8,9,10$\\
			11.1&$(8.5)$&$1,2,3,4,5,6,7,8,9,10,11$\\
			\hline
		\end{tabular}
	\end{table}
\end{landscape}

\newpage
\clearpage
\section{Representations of invariant solutions}\label{AppendixRepInvSol}
Complete results of representations of invariant solutions of Eq. (\ref{eq:fullBoltzannEqWithSources})  are presented in this Appendix. Numbers in the first column are subalgebra numbers of the form \verb|m.n|, where \verb|n| represents number of subalgebra for \verb|m|-dimensional subalgebras. The superscripts $^{c}$, and $^{s}$ which are next to this subalgebra number in the first column indicate that the representation of invariant solution is presented in the cylindrical coordinate system, or spherical coordinate system, respectively. Here $\Omega_{k}$ is an arbitrary function of $k$ independent variables,  and $C$ is constant.

\begin{table}[h]\label{tb:RepresentationsInvariantSolutions}
	\caption{Representations of invariant solutions.}
	\centering
	\begin{tabular}{c|l}
		Subalgebra & Representation of invariant solution $f$ of Eq. (\ref{eq:fullBoltzannEqWithSources}) \\
		\hline
		1.1$^{c}$&$t^{-1}\Omega_{6}(\frac{x}{t}-\frac{\beta}{\alpha}\ln{t},\frac{r}{t},\theta-\frac{1}{\alpha}\ln{t},u-\frac{\beta}{\alpha}\ln{t},V,W)$\\
		1.2$^{c}$&$\Omega_{6}(t,r,\beta\theta-\frac{x}{t},u-\frac{x}{t},V,W)$\\
		1.3$^{c}$&$\Omega_{6}(t,x,r,u,V,W)$\\
		1.4$^{c}$&$\Omega_{6}(t,r,x-\theta,u,V,W)$\\
		1.5$^{c}$&$\Omega_{6}(t^{2}-2x,r,t-\beta\theta,u-t,V,W)$\\
		1.6$^{c}$&$\Omega_{6}(x,r,t-\theta,u,V,W)$\\
		1.7&$t^{-1}\Omega_{6}(\frac{y}{t},\frac{z}{t},\frac{x}{t}-\beta\ln {t},u-\beta\ln {t},v,w)$\\
		1.8&$t^{-1}\Omega_{6}(\frac{x}{t},\frac{y}{t},\frac{z}{t},u,v,w)$\\
		1.9&$\Omega_{6}(t^{2}-2x,y,z,u-t,v,w)$\\
		1.10&$\Omega_{6}(x,y,z,u,v,w)$\\
		1.11&$\Omega_{6}(t,x-tz,y,u-z,v,w)$\\
		1.12&$\Omega_{6}(t,y,z,u-\frac{x}{t},v,w)$\\
		1.13&$\Omega_{6}(t,y,z,u,v,w)$\\
		2.1$^{c}$&$x^{-1}\Omega_{5}(\frac{r}{x},\alpha\theta-\ln{x},u,V,W)$\\
		2.2$^{c}$&$t^{-1}\Omega_{5}(\frac{r}{t},\frac{x}{t}-\alpha\theta-\beta\ln{t},u-\alpha\theta-\beta\ln{t},V,W)$\\
		2.3$^{c}$&$t^{-1}\Omega_{5}(\frac{r}{t},\alpha\theta-\ln{t},u-\frac{x}{t},V,W)$\\
		2.4$^{c}$&$t^{-1}\Omega_{5}(\frac{r}{t},\alpha\theta-\ln{t},u-\frac{\beta}{\alpha}\ln{t},V,W)$\\
		2.5$^{c}$&$\Omega_{5}(r,x,u,V,W)$\\
		2.6$^{c}$&$\Omega_{5}(r,x-\theta,u,V,W)$\\
		2.7$^{c}$&$\Omega_{5}(r,2(x-\alpha\theta)-t^{2},u-t,V,W)$\\
		2.8$^{c}$&$\Omega_{5}(t,r,u-\frac{x}{t},V,W)$\\
		2.9$^{c}$&$\Omega_{5}(t,r,u-\beta\theta,V,W)$\\
		2.10$^{c}$&$\Omega_{5}(t,r,u-\frac{x}{t}+\frac{\theta}{t},V,W)$\\
		2.11$^{c}$&$\Omega_{5}(r,\theta-t,u-\beta t,V,W)$\\
		2.12&$x^{-1}\Omega_{5}(\frac{y}{x},\frac{z}{x},u,v,w)$\\
		2.13&$t^{-1}\Omega_{5}(\frac{y}{t},\frac{z}{t},u-\frac{x}{t},v,w)$\\
		2.14&$t^{-1}\Omega_{5}(\frac{y}{t}-\alpha\ln{t},\frac{z}{t},u-\frac{x}{t},v-\alpha\ln{t},w)$\\
		2.15&$t^{-1}\Omega_{5}(\frac{z}{t},\frac{y}{t}-\alpha\ln{t},u-\beta\ln{t},v-\alpha\ln{t},w)$\\
		2.16&$t^{-1}\Omega_{5}(\frac{y}{t},\frac{z}{t},u-\beta\ln{t},v,w)$\\
		2.17&$\Omega_{5}(y,z,u,v,w)$\\
		2.18&$\Omega_{5}(2x-t^{2},y,u-t,v,w-\alpha t)$\\
		2.19&$\Omega_{5}(y,z,u-t,v,w)$\\
		2.20&$\Omega_{5}(t,\alpha(ty-\tau z)+\beta(tz-\sigma y)+x(\sigma\tau-t^{2}),u,v+\frac{\beta y-\tau x}{\alpha\tau-\beta t},w+\frac{tx-\alpha y}{\alpha\tau-\beta t})$\\
		2.21&$\Omega_{5}(t,x,u,v-\frac{z+ty}{t^2+1},w+\frac{y-tz}{t^2+1})$\\
		2.22&$\Omega_{5}(t,x,u,v-\frac{y}{t},w-\frac{z}{t})$\\

		\hline
	\end{tabular}
\end{table}

\newpage
\begin{table}[h]
	\addtocounter{table}{-1}
	\caption{Representations of invariant solutions (Continued).}
	\centering
	\begin{tabular}{c|l}
		Subalgebra & Representation of invariant solution $f$ of Eq. (\ref{eq:fullBoltzannEqWithSources})\\
		\hline
		2.23&$\Omega_{5}(t,x-\alpha y-tz,u-z,v,w)$\\
		2.24&$\Omega_{5}(t,z,u+\frac{\alpha y-x}{t},v,w)$\\
		2.25&$\Omega_{5}(t,y,u-z,v,w)$\\
		2.26&$\Omega_{5}(t,y,z,v,w)$\\
		2.27&$\Omega_{5}(t,x,u,v,w)$\\
		3.1$^{c}$&$x^{-1}\Omega_{4}(\frac{r}{x},u,V,W)$\\
		3.2$^{c}$&$e^{-\alpha\theta}\Omega_{4}(re^{-\alpha\theta},u-\beta\theta,V,W)$\\
		3.3$^{c}$&$t^{-1}\Omega_{4}(\frac{r}{t},u-\frac{x}{t},V,W)$\\
		3.4$^{c}$&$t^{-1}\Omega_{4}(\frac{r}{t},u-\alpha\theta-\beta\ln{t},V,W)$\\
		3.5&$t^{-1}\Omega_{4}(\frac{x}{t}-\frac{\beta}{\alpha}\ln{t},u-\frac{\beta}{\alpha}\ln{t},\arctan(\frac{w-\frac{z}{t}}{v-\frac{y}{t}})-\frac{1}{\alpha}\ln{t},\sqrt{(v-\frac{y}{t})^{2}+(w-\frac{z}{t})^{2}})$\\
		3.6$^{c}$&$t^{-1}\Omega_{4}(\frac{r}{t},\alpha\theta-\ln{t},V,W)$\\
		3.7&$t^{-1}\Omega_{4}(\frac{x}{t}-\frac{\beta}{\alpha}\ln{t},u-\frac{\beta}{\alpha}\ln{t},\arctan(\frac{w}{v})-\frac{1}{\alpha}\ln{t},\sqrt{v^{2}+w^{2}})$\\
		3.8$^{s}$&$\Omega_{4}(t,r,U,V^{2}+W^{2})$\\
		3.9$^{c}$&$\Omega_{4}(r,u-t-\alpha\theta,V,W)$\\
		3.10&$\Omega_{4}(t,\frac{x}{t}-\beta\arctan(\frac{w-\frac{z}{t}}{v-\frac{y}{t}}),u-\beta\arctan(\frac{w-\frac{z}{t}}{v-\frac{y}{t}}),\sqrt{(v-\frac{y}{t})^{2}+(w-\frac{z}{t})^{2}})$\\
		3.11$^{c}$&$\Omega_{4}(t,r,V,W)$\\
		3.12&$\Omega_{4}(t,u-\frac{x}{t},\arctan(\frac{w}{v})-\frac{x}{\beta t},\sqrt{v^{2}+w^{2}})$\\
		3.13&$\Omega_{4}(t,x,u,\sqrt{v^{2}+w^{2}})$\\
		3.14&$\Omega_{4}(t,u-\frac{\alpha x}{1+\alpha t},\arctan(\frac{w-\frac{z}{t}}{v-\frac{y}{t}})-\frac{x}{1+\alpha t},\sqrt{(v-\frac{y}{t})^{2}+(w-\frac{z}{t})^{2}})$\\
		3.15&$\Omega_{4}(t,u+\beta\arctan(\frac{y-tv+w}{z-tw-v}),x+(\alpha+\beta t)\arctan(\frac{y-tv+w}{z-tw-v}),$\\
		&$\sqrt{(y-tv+w)^{2}+(z-tw-v)^{2}})$\\
		3.16&$\Omega_{4}(t,u,\arctan(\frac{w}{v})-x,\sqrt{v^{2}+w^{2}})$\\
		3.17$^{c}$&$\Omega_{4}(t-\theta,r,V,W)$\\
		3.18&$\Omega_{4}(2x-t^{2},u-t,\arctan(\frac{w}{v})-\frac{t}{\beta},\sqrt{v^{2}+w^{2}})$\\
		3.19&$\Omega_{4}(x,u,\arctan(\frac{w}{v})-t,\sqrt{v^{2}+w^{2}})$\\
		3.20&$y^{-1}\Omega_{4}(\frac{y}{z},u-\beta\ln{y},v,w)$\\
		3.21&$t^{-1}\Omega_{4}(\frac{x}{t}-\beta\ln{t},u-\beta\ln{t},v-\frac{y}{t},w-\frac{z}{t})$\\
		3.22&$t^{-1}\Omega_{4}(\frac{y}{t}-\sigma\ln{t},u-\frac{\alpha z}{t}-\beta\ln{t},v-\sigma\ln{t},w-\frac{z}{t})$\\
		3.23&$t^{-1}\Omega_{4}(\frac{y}{t},\frac{z}{t}-\sigma\ln{t},v,w-\sigma\ln{t})$\\
		3.24&$t^{-1}\Omega_{4}(\frac{y}{t},\frac{z}{t},v,w)$\\
		3.25&$t^{-1}\Omega_{4}(\frac{x}{t}-\beta\ln{t},u-\beta\ln{t},v-\sigma\ln{t},w)$\\
		3.26&$t^{-1}\Omega_{4}(\frac{x}{t}-\beta\ln{t},u-\beta\ln{t},v,w)$\\
		3.27&$\Omega_{4}(y-\frac{\beta x}{\alpha}+\frac{\beta t^{2}}{2\alpha},u-t,v,w-\frac{x}{\alpha}+\frac{t^{2}}{2\alpha}),~\alpha\neq0$\\
		&$\Omega_{4}(t^{2}-2x,u-t,v,w-\frac{y}{\beta}),~\alpha=0,~\beta\neq0$\\
		&$\Omega_{4}(t^{2}-2x,y,u-t,v),~\alpha=0,~\beta=0$\\
		3.28&$\Omega_{4}(z,u-y,v,w)$\\
		3.29&$\Omega_{4}(y,z,v,w)$\\
		3.30&$\Omega_{4}(t^{2}-2x,u-t,v,w-\sigma t)$\\
		3.31&$\Omega_{4}(t^{2}-2x,u-t,v,w)$\\
		3.32&$\Omega_{4}(x,u,v,w-t)$\\
		3.33&$\Omega_{4}(x,u,v,w)$\\
		3.34&$\Omega_{4}(t,x-tu-\delta v+\beta w,y-tv+\delta u-\sigma v-\alpha w,z-tw-\beta u+\alpha v-\tau w)$\\
		\hline
	\end{tabular}
\end{table}

\newpage
\begin{table}[h]
	\addtocounter{table}{-1}
	\caption{Representations of invariant solutions (Continued).}
	\centering
	\begin{tabular}{c|l}
		Subalgebra  & Representation of invariant solution $f$ of Eq. (\ref{eq:fullBoltzannEqWithSources})\\
		\hline

		3.35&$\Omega_{4}(u-\frac{x}{t},v-\frac{y}{t}-\frac{tz-y}{t(t^{2}+1)},w-\frac{tz-y}{t^{2}+1},t)$\\
		3.36&$\Omega_{4}(u-\frac{x}{t+1},v-\frac{y}{t},w-\frac{z}{t},t)$\\
		3.37&$\Omega_{4}(u-\frac{x}{t},v-\frac{y}{t},w-\frac{z}{t},t)$\\
		3.38&$\Omega_{4}(u,v-\frac{y}{t}+\frac{\tau(x-\alpha z-\frac{\beta y}{t})}{\alpha t^{2}-\sigma t+\tau\beta},w+\frac{t(x-\alpha z-\frac{\beta y}{t})}{\alpha t^{2}-\sigma t+\tau\beta},t)$\\
		3.39&$\Omega_{4}(u,v-\frac{y}{t},w+\frac{x-\alpha z}{\alpha t},t),\alpha\neq0$\\
		&$\Omega_{4}(x,u,v-\frac{y}{t},t),~\alpha=0$\\
		3.40&$\Omega_{4}(t,w(t^{2}-\tau)+y-tz,\tau w-y+tv,u)$\\
		3.41&$\Omega_{4}(t,w(t^{2}+1)+y-tz,w+y-tv,u)$\\
		3.42&$\Omega_{4}(t,u,v-\frac{y}{t},w-\frac{z}{t})$\\
		3.43&$\Omega_{4}(t,tu-x+\beta z,v,w)$\\
		3.44&$\Omega_{4}(t,u-\frac{x}{t},v,w)$\\
		3.45&$\Omega_{4}(t,u-z,v,w)$\\
		3.46&$\Omega_{4}(t,z,v,w)$\\
		3.47&$\Omega_{4}(t,u,v,w)$\\
		4.1$^{s}$&$r^{-1}\Omega_{3}(\frac{t}{r},U,V^{2}+W^{2})$\\
		4.2$^{c}$&$r^{-1}\Omega_{3}(u-\alpha\theta,V,W)$\\
		4.3&$e^{\alpha\arctan(\frac{v}{w})}\Omega_{3}(xe^{\alpha\arctan(\frac{v}{w})},\sqrt{v^{2}+w^{2}},u)$\\
		4.4$^{c}$&$r^{-1}\Omega_{3}(\alpha\theta-\ln{r},V,W)$\\
		4.5&$t^{-1}\Omega_{3}(\frac{\sqrt{(y-tv)^{2}+(z-tw)^{2}}}{t},t^{-\beta}e^{\frac{x-\alpha t \arctan(\frac{z-tw}{y-tv})}{t}},t^{-\beta}e^{u-\alpha\arctan(\frac{z-tw}{y-tv})})$\\
		4.6$^{c}$&$r^{-1}\Omega_{3}(\frac{t}{r},V,W)$\\
		4.7&$t^{-1}\Omega_{3}(t^{-\beta}e^{\frac{x+\alpha t\arctan(\frac{v}{w})}{t}},t^{-\beta}e^{u+\alpha\arctan(\frac{v}{w})},\sqrt{v^{2}+w^{2}})$\\
		4.8&$t^{-1}\Omega_{3}(\frac{x-tu}{t},\frac{\sqrt{(y-tv)^{2}+(z-tw)^{2}}}{t},te^{-\alpha\arctan(\frac{z-tw}{y-tv})})$\\
		4.9&$e^{-\alpha\arctan(\frac{z-tw}{y-tv})}\Omega_{3}(\sqrt{(y-tv)^{2}+(z-tw)^{2}}e^{-\alpha\arctan(\frac{z-tw}{y-tv})},te^{-\alpha\arctan(\frac{z-tw}{y-tv})},$\\
		&$u-\beta\arctan(\frac{z-tw}{y-tv}))$\\
		4.10&$e^{\alpha\arctan(\frac{v}{w})}\Omega_{3}((x-tu)e^{\alpha\arctan(\frac{v}{w})},te^{\alpha\arctan(\frac{v}{w})},\sqrt{v^{2}+w^{2}})$\\
		4.11&$t^{-1}\Omega_{3}(te^{\alpha\arctan(\frac{v}{w})},t^{-\beta}e^{\alpha u},\sqrt{v^{2}+w^{2}})$\\
		4.12&$\Omega_{3}(u+\beta\arctan(\frac{v}{w}),\sqrt{v^{2}+w^{2}},t)$\\
		4.13$^{s}$&$\Omega_{3}(r,U,V^{2}+W^{2})$\\
		4.14&$\Omega_{3}(x,u,v^{2}+w^{2})$\\
		4.15&$\Omega_{3}(x+\arctan(\frac{v}{w}),\sqrt{v^{2}+w^{2}},u)$\\
		4.16&$\Omega_{3}(2\alpha\arctan(\frac{v}{w})+2x-t^{2},\sqrt{v^{2}+w^{2}},u-t)$\\
		4.17&$\Omega_{3}(u-\frac{x}{t},(v-\frac{y}{t})^{2}+(w-\frac{z}{t})^{2},t)$\\
		4.18&$\Omega_{3}(x-tu-\arctan(\frac{z-tw}{y-tv}),\sqrt{(y-tv)^{2}+(z-tw)^{2}},t)$\\
		4.19&$\Omega_{3}(x-tu-\alpha\arctan(\frac{z-v-tw}{w+y-tv}),\sqrt{(w+y-tv)^{2}+(z-v-tw)^{2}},t)$\\
		4.20&$\Omega_{3}(u-\alpha\arctan(\frac{z-v-tw}{w+y-tv}),\sqrt{(w+y-tv)^{2}+(z-v-tw)^{2}},t)$\\
		4.21&$\Omega_{3}(x-tu+\arctan(\frac{v}{w}),\sqrt{v^{2}+w^{2}},t)$\\
		4.22&$\Omega_{3}(u-\beta t,t+\arctan(\frac{v}{w}),\sqrt{v^{2}+w^{2}})$\\
		4.23&$y^{-1}\Omega_{3}(\frac{z}{y},v,w)$\\
		4.24&$x^{-1}\Omega_{3}(u,v,w-\alpha\ln{x})$\\
		4.25&$x^{-1}\Omega_{3}(u,v,w)$\\
		4.26&$t^{-1}\Omega_{3}(u-\frac{x}{t},v-\frac{y}{t},w-\frac{z}{t})$\\
		\hline
	\end{tabular}
\end{table}

\newpage
\begin{table}[h]
	\addtocounter{table}{-1}
	\caption{Representations of invariant solutions (Continued).}
	\centering
	\begin{tabular}{c|l}
		Subalgebra & Representation of invariant solution $f$ of Eq. (\ref{eq:fullBoltzannEqWithSources})\\
		\hline

		4.27&$t^{-1}\Omega_{3}(\frac{y-tv}{t},\frac{z-tw}{t},t^{-\beta}e^{u-\alpha v})$\\
		4.28&$t^{-1}\Omega_{3}(u-\beta\ln{t},v-\frac{y}{t},w-\frac{z}{t})$\\
		4.29&$t^{-1}\Omega_{3}(\frac{y}{t}-\alpha\ln{t},v-\alpha\ln{t},w-\frac{z}{t})$\\
		4.30&$t^{-1}\Omega_{3}(t^{-\beta}e^{\frac{x-\alpha tw}{t}},t^{-\beta}e^{u-\alpha w},t^{-\sigma}e^{v})$\\
		4.31&$t^{-1}\Omega_{3}(u-\frac{x}{t},v-\alpha\ln{t},w-\beta\ln{t})$\\
		4.32&$t^{-1}\Omega_{3}(u-\frac{x}{t},v,w)$\\
		4.33&$t^{-1}\Omega_{3}(u-\beta\ln{t},v,w)$\\
		4.34&$t^{-1}\Omega_{3}(u,v,w)$\\
		4.35&$\Omega_{3}(u-t,v-\frac{x}{\alpha}+\frac{t^{2}}{2\alpha},w-\beta t),~\alpha\neq0$\\
		&$\Omega_{3}(t^{2}-2x,u-t,w-\beta t),~\alpha=0$\\
		4.36&$\Omega_{3}(u,v-\frac{x}{\alpha},w-t),~\alpha\neq0$\\
		&$\Omega_{3}(x,u,w-t),~\alpha=0$\\
		4.37&$\Omega_{3}(u,v-x,w)$\\
		4.38&$\Omega_{3}(x,u,w)$\\
		4.39&$\Omega_{3}(u-t,v,w)$\\
		4.40&$\Omega_{3}(u,v,w)$\\
		4.41&$\Omega_{3}(y-\sigma u-tv-\beta w,z-\tau u-\alpha v-tw,t)$\\
		4.42&$\Omega_{3}(y-tv+w,y-tz+w(t^{2}+1),t)$\\
		4.43&$\Omega_{3}(v-\frac{y}{t},w-\frac{z}{t},t)$\\
		4.44&$\Omega_{3}(v-x+\alpha (z-tw),u,t)$\\
		4.45&$\Omega_{3}(u,v-x,t)$\\
		4.46&$\Omega_{3}(t,u,w+\frac{\beta x-z}{t})$\\
		4.47&$\Omega_{3}(x,u,t)$\\
		4.48&$\Omega_{3}(v-z+tw,u,t)$\\
		4.49&$\Omega_{3}(u,w-\frac{z}{t},t)$\\
		4.50&$\Omega_{3}(v,w,t)$\\
		5.1$^{s}$&$r^{-1}\Omega_{2}(U,V^{2}+W^{2})$\\
		5.2$^{c}$&$r^{-1}\Omega_{2}(V,W)$\\
		5.3&$x^{-1}\Omega_{2}(v^{2}+w^{2},u)$\\
		5.4&$e^{\alpha\arctan(\frac{v}{w})}\Omega_{2}(u+\beta\arctan(\frac{v}{w}),\sqrt{v^{2}+w^{2}})$\\
		5.5&$t^{-1}\Omega_{2}(\frac{x-tu}{t},\frac{(y-tv)^{2}+(z-tw)^{2}}{t^{2}})$\\
		5.6&$t^{-1}\Omega_{2}(\frac{x-tu}{t},v^{2}+w^{2})$\\
		5.7&$t^{-1}\Omega_{2}(t^{\beta}e^{\alpha\arctan(\frac{z-tw}{y-tv})-u},\sqrt{(\frac{y-tv}{t})^{2}+(\frac{z-tw}{t})^{2}})$\\
		5.8&$t^{-1}\Omega_{2}(t^{-\beta}e^{u+\alpha\arctan(\frac{v}{w})},\sqrt{v^{2}+w^{2}})$\\
		5.9&$e^{-\alpha\arctan(\frac{z-tw}{y-tv})}\Omega_{2}(\sqrt{(y-tv)^{2}+(z-tw)^{2}}e^{-\alpha\arctan(\frac{z-tw}{y-tv})},te^{-\alpha\arctan(\frac{z-tw}{y-tv})})$\\
		5.10&$t^{-1}\Omega_{2}(\frac{x}{t}-\frac{\beta}{\alpha}\ln{t},u-\frac{\beta}{\alpha}\ln{t})$\\
		5.11&$t^{-1}\Omega_{2}(te^{\alpha\arctan(\frac{v}{w})},\sqrt{v^{2}+w^{2}})$\\
		5.12&$\Omega_{2}(t-u-\alpha\arctan(\frac{v}{w}),\sqrt{v^{2}+w^{2}})$\\
		5.13&$\Omega_{2}(u-\frac{x}{t},t)$\\
		5.14&$(4.47)$\\
		5.15&$\Omega_{2}(v^{2}+w^{2},t)$\\
		5.16&$\Omega_{2}((w+y-tv)^{2}+(v-z+tw)^{2},t)$\\
		\hline
	\end{tabular}
\end{table}

\newpage
\begin{table}
	\addtocounter{table}{-1}
	\caption{Representations of invariant solutions (Continued).}
	\centering
	\begin{tabular}{c|l}
		Subalgebra & Representation of invariant solution $f$ of Eq. (\ref{eq:fullBoltzannEqWithSources})\\
		\hline

		5.17&$\Omega_{2}(u,t)$\\
		5.18&$\Omega_{2}(t^{2}-2x,u-t)$\\
		5.19&$\Omega_{2}(x,u)$\\
		5.20&$\Omega_{2}(t+\arctan(\frac{v}{w}),\sqrt{v^{2}+w^{2}})$\\
		5.21&$x^{-1}\Omega_{2}(u,w-\beta\ln{x})$\\
		5.22&$e^{-\beta u}\Omega_{2}(v,w)$\\
		5.23&$None$\\
		5.24&$t^{-1}\Omega_{2}(v-\frac{y}{t},w-\frac{z}{t})$\\
		5.25&$t^{-1}\Omega_{2}(u-\frac{x}{t},v-\frac{x}{\alpha t}+\frac{\beta}{\alpha}\ln{t})$\\
		5.26&$t^{-1}\Omega_{2}(\frac{x}{t}-\beta\ln{t},u-\beta\ln{t})$\\
		5.27&$t^{-1}\Omega_{2}(u-\frac{x}{t},v-\beta\ln{t})$\\
		5.28&$t^{-1}\Omega_{2}(u-\beta\ln{t},v)$\\
		5.29&$t^{-1}\Omega_{2}(v,w)$\\
		5.30&$\Omega_{2}(2(x-\alpha v)-t^{2},u-t)$\\
		5.31&$\Omega_{2}(t^{2}-2x,u-t)$\\
		5.32&$\Omega_{2}(u,v-x)$\\
		5.33&$\Omega_{2}(x,u)$\\
		5.34&$\Omega_{2}(u-t,v)$\\
		5.35&$\Omega_{2}(u-\frac{x}{t},t)$\\
		5.36&$\Omega_{2}(t,tu-x+w)$\\
		5.37&$\Omega_{2}(u,t)$\\
		6.1&$None$\\
		6.2&$x^{-1}\Omega_{1}(u)$\\
		6.3&$e^{\alpha\arctan(\frac{v}{w})}\Omega_{1}(\sqrt{v^{2}+w^{2}})$\\
		6.4&$t^{-1}\Omega_{1}((v-\frac{y}{t})^{2}+(w-\frac{z}{t})^{2})$\\
		6.5&$t^{-1}\Omega_{1}(v^{2}+w^{2})$\\
		6.6&$t^{-1}\Omega_{1}(u-\frac{x}{t}),~\alpha\neq0$\\
		&$(5.26),~\alpha=0$\\
		6.7&$t^{-1}\Omega_{1}(u-\frac{x}{t})$\\
		6.8&$e^{-\frac{\alpha u}{\beta}}\Omega_{1}(te^{-\frac{\alpha u}{\beta}}),~\beta\neq0$\\
		&$t^{-1}\Omega_{1}(u),~\beta=0,~\alpha\neq0$\\
		&$(5.37),~\beta=0,~\alpha=0$\\
		6.9&$\Omega_{2}(\sqrt{(u-\frac{x}{t})^{2}+(v-\frac{y}{t})^{2}+(w-\frac{z}{t})^{2}},t)$\\
		6.10&$\Omega_{2}(\sqrt{u^{2}+v^{2}+w^{2}},t)$\\
		6.11&$\Omega_{1}(u)$\\
		6.12&$\Omega_{1}(u-t),~\alpha\neq0$\\
		&$(5.31),~\alpha=0$\\
		6.13&$(5.33)$\\
		6.14&$\Omega_{1}(t)$\\
		6.15&$(5.35)$\\
		6.16&$\Omega_{1}(u)$\\
		6.17&$x^{-1}\Omega_{1}(u)$\\
		6.18&$e^{-\frac{w}{\alpha}}\Omega_{1}(v)$\\
		\hline
	\end{tabular}
\end{table}

\newpage
\begin{table}[h]
	\addtocounter{table}{-1}
	\caption{Representations of invariant solutions (Continued).}
	\centering
	\begin{tabular}{c|l}
		Subalgebra & Representation of invariant solution $f$ of Eq. (\ref{eq:fullBoltzannEqWithSources})\\
		\hline
		6.19&$None$\\
		6.20&$t^{-1}\Omega_{1}(u-\alpha\ln{t})$\\
		6.21&$t^{-1}\Omega_{1}(u-\frac{x}{t})$\\
		6.22&$\Omega_{1}(u)$\\
		6.23&$\Omega_{1}(u-t)$\\
		6.24&$\Omega_{1}(t)$\\
		6.25&$\Omega_{1}(t),~\beta\neq0$\\
		&$(5.37),~\beta=0$\\
		7.1&$t^{-1}\Omega_{1}(\sqrt{u^{2}+v^{2}+w^{2}})$\\
		7.2&$t^{-1}\Omega_{1}(\sqrt{(u-\frac{x}{t})^{2}+(v-\frac{y}{t})^{2}+(w-\frac{z}{t})^{2}})$\\
		7.3&$(6.17)$\\
		7.4&$None$\\
		7.5&$Ce^{-\frac{\alpha u}{\beta}},~\beta\neq0$\\
		&$None,~\beta=0,~\alpha\neq0$\\
		&$(6.22),~\beta=0,~\alpha=0$\\
		7.6&$(6.21)$\\
		7.7&$Ct^{-1},~\alpha\neq0$\\
		&$(6.20),~\alpha=0$\\
		7.8&$Ct^{-1}$\\
		7.9&$\Omega_{1}(\sqrt{u^{2}+v^{2}+w^{2}})$\\
		7.10&$C,~\alpha\neq0$\\
		&$(6.23),~\alpha=0$\\
		7.11&$C$\\
		7.12&$Ce^{-\frac{u}{\alpha}}$\\
		7.13&$Ct^{-1}$\\
		7.14&$C$\\
		8.1&$None$\\
		8.2&$None,~\alpha\neq0$\\
		&$(7.12),~\alpha=0$\\
		8.3&$None,~\alpha\neq0$\\
		&$(7.14),~\alpha=0$\\
		8.4&$(7.13)$\\
		8.5&$None$\\
		9.1&$(8.5)$\\
		9.2&$(6.24)$\\
		10.1&$(7.13)$\\
		10.2&$(7.14)$\\
		11.1&$(8.5)$\\
		\hline
	\end{tabular}
\end{table}




\newpage
\clearpage
\section*{References}



\end{document}